\documentclass[twocolumn]{aastex61}

\usepackage{hyperref}
\usepackage{xspace}
\usepackage{graphicx}
\usepackage{verbatim}

\usepackage{bm}
\usepackage{amssymb}
\usepackage{epstopdf}
\usepackage{amsmath}
\usepackage{amsfonts}
\usepackage{mathrsfs}
\usepackage{needspace}

\newcommand{\refchange}[1]{#1}
\newcommand{\refchangetwo}[1]{#1}

\usepackage{amsmath}

\renewcommand{\la}{\ensuremath{\left\langle}}  
\newcommand{\ra}{\ensuremath{\right\rangle}}

\renewcommand{\vec}[1]{{\boldsymbol{\mathbf{#1}}}}

\newcommand{\vu}{\vec{u}}

\newcommand{\vx}{\vec{x}}

\newcommand{\vnhat}{\hat{\vec{n}}}

\newcommand{\vxhat}{\hat{\vec{x}}}

\newcommand{\vE}{\vec{E}}

\newcommand{\veps}{\vec{\varepsilon}}

\DeclareMathOperator{\sinc}{sinc}

\DeclareMathOperator{\Cov}{Cov}

\newcommand{\calP}{\mathcal{P}}

\newcommand{\dhn}{d^2\hat{\vec{n}}}

\newcommand{\eps}{\varepsilon}

\newcommand{\nant}{n_{\rm ant}}

\newcommand{\be}{\begin{equation}}
\newcommand{\ee}{\end{equation}}
\newcommand{\ba}{\begin{eqnarray}}
\newcommand{\ea}{\end{eqnarray}}
\newcommand{\nn}{\nonumber}
\newcommand{\barr}{\begin{array}}
\newcommand{\earr}{\end{array}}

\newcommand\lsim{\mathrel{\rlap{\lower4pt\hbox{\hskip1pt$\sim$}}
        \raise1pt\hbox{$<$}}}
\newcommand\gsim{\mathrel{\rlap{\lower4pt\hbox{\hskip1pt$\sim$}}
        \raise1pt\hbox{$>$}}}

\def\n{{\bf n}}

\def\Var{\mbox{Var}}

\newcommand{\aindex}{a}
\newcommand{\bindex}{b}
\newcommand{\cindex}{c}
\newcommand{\dindex}{d}
\newcommand{\jindex}{j}
\newcommand{\kindex}{k}
\newcommand{\alphaindex}{\alpha}
\newcommand{\deltaindex}{\delta}

\newcommand{\sumalpha}{%
    \sum_\alphaindex
    }
\newcommand{\sumdelta}{%
    \sum_\deltaindex
    }

\renewcommand{\eqref}[1]{Equation~\ref{#1}}

\begin{document}

\title{Algorithms for FFT Beamforming Radio Interferometers}

\correspondingauthor{Kiyoshi~W.~Masui}
\email{kmasui@mit.edu}

\author{Kiyoshi~W.~Masui}
\affiliation{MIT Kavli Institute for Astrophysics and Space Research,
Massachusetts Institute of Technology,
77 Massachusetts Avenue, Cambridge, MA~02139, United States}
\affiliation{Department of Physics,
Massachusetts Institute of Technology,
77 Massachusetts Avenue, Cambridge, MA~02139, United States}

\author{J.~Richard Shaw}
\affiliation{Department of Physics and Astronomy, University of British
        Columbia, 6224 Agricultural Rd, Vancouver, BC~V6T~1Z1, Canada}

\author{Cherry Ng}
\affiliation{Dunlap Institute, University of Toronto,
    50 St. George St., Toronto, ON~M5S~3H4, Canada}

\author{Kendrick~M.~Smith}
\affiliation{Perimeter Institute for Theoretical Physics, Waterloo,
    ON N2L 2Y5, Canada}

\author{Keith Vanderlinde}
\affiliation{Dunlap Institute, University of Toronto,
    50 St. George St., Toronto, ON~M5S~3H4, Canada}
\affiliation{Department of Astronomy and Astrophysics, University of Toronto,
    50 St George St., Toronto, ON~M5S~3H4, Canada}

\author{Adiv Paradise}
\affiliation{Department of Astronomy and Astrophysics, University of Toronto,
    50 St George St., Toronto, ON~M5S~3H4, Canada}

\date{\today}


\begin{abstract}
Radio interferometers consisting of identical antennas arranged on a regular
lattice permit fast Fourier transform beamforming, which reduces the correlation
cost from $\mathcal{O}(n^2)$ in the number of antennas to
$\mathcal{O}(n\log n)$. We
develop a formalism for describing this process and apply this formalism to
derive a number of algorithms with a range of observational applications.
These include algorithms for
forming arbitrarily pointed tied-array beams from the regularly spaced
Fourier-transform formed beams, sculpting the beams to suppress sidelobes
while only losing percent-level sensitivity, and optimally estimating the
position of a detected source from its observed brightness in the set of beams.
We also discuss the effect that correlations in the visibility-space noise, due to
cross-talk and sky contributions, have on the optimality of Fourier transform
beamforming, showing that it does not strictly preserve the sky information of
the $n^2$ correlation, even for an idealized array. Our results have
\refchange{applications to} a number of upcoming interferometers, in particular the
Canadian Hydrogen Intensity Mapping Experiment--Fast Radio Burst
(CHIME/FRB) project.
\end{abstract}



\section{Introduction}

Interferometry has been central to the field of radio astronomy for 70
years. Interferometers combine the signals from multiple antennas coherently
to both increase sensitivity and gain spatial information. Many of today's most
successful radio observatories are interferometers with many dozen
antennas.

In the past, the size of interferometers has been limited by the cost of the
electronics that instrument the antennas and the computational cost to combine
their signals. However, the latter \refchange{has become}
less challenging with Moore's Law and
the former has become dramatically cheaper with the advent of mass-produced
electronics designed for the communications industry. This has permitted a new
class of radio telescope composed of a large number---hundreds to thousands---of
low-cost, typically non-steerable, antennas. These include 
CHIME\footnote{\url{https://chime-experiment.ca}} \citep{2014SPIE.9145E..22B},
HERA\footnote{\url{http://reionization.org}} \citep{2017PASP..129d5001D},
HIRAX \citep{2016SPIE.9906E..5XN},
LEDA\footnote{\url{http://www.tauceti.caltech.edu/leda}} \citep{2012AAS...22010403G,2017arXiv170909313P},
LOFAR\footnote{\url{http://lofar.org}} \citep{2013A&A...556A...2V},
MITEoR \citep{2014MNRAS.445.1084Z},
MWA\footnote{\url{http://mwatelescope.org}} \citep{2009IEEEP..97.1497L},
the Ooty Radio Telescope \citep{2013arXiv1310.1707S},
\mbox{PAPER}\footnote{\url{http://eor.berkeley.edu}},
Tianlai\footnote{\url{http://tianlai.bao.ac.cn}} \citep{2012IJMPS..12..256C},
and
UTMOST\footnote{\url{https://astronomy.swin.edu.au/research/utmost}} \citep{2016MNRAS.458..718C}
.

\refchange{Further scaling of this type of instrument is limited by the}
computational cost to pairwise correlate the
antenna signals, which scales as $n^2$ in the number of antennas compared to $n$ for
the
mechanical and analogue components of the telescope. As such, beyond a certain
number of antennas, the telescope cost will once again be dominated by the
computational correlation cost, even while the cost of computation is dropping over time.
An alternate form of correlation was used on the Waseda Radio Telescope
\citep{1992PASJ...44L..35N, 1994PASJ...46..503O, daishido2000}
using a fast Fourier transform (FFT) of the antenna
signals in the spatial (antenna-position) direction rather
than pairwise correlation. The output of this process is localized beams
on the sky rather than visibilities.
In \citet{2004NewA....9..417P} it was suggested that this method could
be used for very large interferometers to reduce the correlation cost
to scale as $n\log n$, an idea that was formalized and extended in
\citet{2009PhRvD..79h3530T, 2010PhRvD..82j3501T} and implemented in on the BEST-2
array by \citet{2014MNRAS.439.3180F}.
\refchange{
The concept was further extended to apply to irregular and heterogeneous arrays of antennas by
\citet{2011PASP..123.1265M}, which was extended and implemented in \citet{2017MNRAS.467..715T} and
\citet{2017MNRAS.470.4720B}.
}
\refchange{FFT beamforming
dramatically reduces the correlation cost, which in principle should
allow for the construction of
telescopes with many more antennas that will be orders
of magnitude more sensitive than current instruments.}
It is envisaged that such telescopes will permit
neutral hydrogen gas to be mapped over large volumes of the high-redshift
Universe, spurring a revolution in observational cosmology
\citep{2004PhRvL..92u1301L,2006PhR...433..181F,2010PhRvL.105p1302M, 2011PASP..123.1265M}.

In the near term, FFT beamforming will be used at the CHIME (specifically CHIME/FRB,
\citealp{2018ApJ...863...48C, 2017arXiv170204728N}) and HIRAX telescopes to
\refchange{correlate roughly
two thousand antenna signals and}
search for fast radio bursts. In this
application, the calibration challenges that currently prevent FFT
beamforming from being
used for hydrogen surveys \citep{2009MNRAS.394.1575L, 2010MNRAS.408.1029L,
2014SPIE.9145E..4VN} are less severe.
\refchange{In hydrogen surveys the foregrounds are several orders of magnitude
brighter than the signal, so small calibration errors can lead to a small fraction
of the foregrounds leaking into the signal channel which then swamps the signal.
When using FFT beamforming calibration must be performed in real time, whereas
in traditional correation it can be done in offline analysis,
allowing for a more careful calibration.
On the other hand, FRBs are separated from contaminants in the time-domain,
and as such the main concern
is sensitivity of the telescope to sky signals. We will discuss the effect of
calibration errors in more detail in Section~\ref{s:nonredundancy}.
}

The use of FFT beamforming in FRB searches
does present other challenges however. The simplest FFT beamforming
algorithms give little control over the locations of the beams on the sky, and
these locations are wavelength dependant. Transient surveys typically need to
maximize instantaneous broadband sensitivity to a single location, rather than form a map of
the static sky. As such, the chromaticity of the beam locations
must be dealt with in some way, but the
simplest methods of doing so introduce severe spectral structure in the beam
shape \citep{2017arXiv170204728N}. Another issue is a poor understanding of the
noise properties of individual beams and how it is correlated between them. This
has led to confusion in how well a transient source can be localized from a
multi-beam detection, and the optimal algorithm for doing so.

In this article, we develop a formalism for beamforming, particularly focusing
on FFT beamforming. We use this to address the issues discussed above and derive
a number of algorithms with a range of observational applications.
To orient the reader, the highlights of our work are summarized as follows.
The formed beam that optimizes its response to a single point on the sky is given
in Equation~\ref{e:wp} or Equation~\ref{e:wopt} depending on the generality of the
noise model assumed. In Section~\ref{s:fft_beamforming} we show that, for
redundant arrays, using an FFT to form $2\nant -1$ beams has the same information
content as the visibilities, but only if simplifying assumptions are made about
the noise.
For forming a large number of beams (for example ``fan beams'' to
perform blind searches
for sources),
Equation~\ref{e:bp_max_arbitrary} allows the FFT beams to be exactly regridded
to arbitrary (and achromatic) positions using downsampled intensities.
Section~\ref{s:windowed_beams} describes how a form of windowing can
be used to suppress sidelobes, decrease the regridding cost, and increase
beam solid angle while losing only a small amount of peak sensitivity. This
results in a net higher discovery rate in blind searches when the number of
beams that can be searched is fixed.
In Section~\ref{s:localization} we derive the optimal estimator for the
location of a source from a multi-beam detection.
In a follow-up
work, we will use the strategies described here to perform a comprehensive
optimization for upcoming experiments like CHIME/FRB.


\section{Preliminaries}

In this section we will introduce our notation and conventions for describing
the sky and instrument response. Our notation is based on that developed in
\citet{2014ApJ...781...57S, 2015PhRvD..91h3514S},
\refchange{although the underlying
derivations are presented in many other works 
\citep{1934Phy.....1..201V,1938Phy.....5..785Z,1996A&AS..117..137H,2011A&A...527A.106S,2017isra.book.....T}.}
One source of complexity in
this work is the large number of different types of indices used to iterate
over different spaces. For clarity, we summarize these in Table~\ref{t:inds}.

\begin{table*}
\newlength{\valsep}
\setlength{\valsep}{0.8mm}
\begin{center}
\begin{tabular}{l p{0.4\linewidth} l}
Symbol & Quantity indexed & Range\\[\valsep]
\hline
$\jindex$, $\kindex$, $l$, $m$ & Directions perpendicular to incident radiation & 0 to 1\\[\valsep]
    $P$ & The Stokes parameters & $( I, Q, U, V)$\\[\valsep]
$\aindex$, $\bindex$, $\cindex$, $\dindex$ & Antennas & 0 to $\nant -1$ \\[\valsep]
$\delta$ & Difference between two feed indices $\aindex - \bindex$ & $-(\nant - 1)$ to $\nant - 1$\\[\valsep]
$\alpha$ & ``Redundancy'' index complementary to $\delta$. 
    & $\max(0, -\delta)$ to $\min(\nant ,\nant - \delta)$\\[\valsep]
$A$, $B$, $C$, $D$ & FFT formed beams & 0 to $M - 1$\\[\valsep]
    \refchange{$X$, $Y$} & \refchange{Label for beams within a generic set} &
    \refchange{The full set}\\[\valsep]
\hline
\end{tabular}
\end{center}
    \caption{\label{t:inds} Indices used, their meaning, and implied
    summation limits unless otherwise given.}
\end{table*}

\subsection{Sky}

An antenna samples the electric field in
a weighted volume surrounding its location. In detail this response is complicated as in
the near-field the antenna itself serves to modify the electric field,
however in our case we only need the far field response. To start we
write the electric field in absence of the antenna as the sum of plane waves
coming from the far field
\begin{equation}
    \vE(\vx, t) = \frac{1}{(\epsilon_0 c)^\frac{1}{2}} \int \veps(\vnhat, \nu)
    e^{-i2\pi \nu (t - \vx\cdot\vnhat / c)} \, \dhn \, d\nu\;,
\end{equation}
which defines the quantity $\veps(\vnhat, \nu)$. \refchange{Here $\vnhat$ is
a unit vector defining a direction on the sky, and $\dhn$ is the differential
solid angle.} That the electric field is
real sets $\veps(\vnhat, \nu) = \veps(\vnhat, -\nu)^*$.

We are generally not interested in the actual phase of the incoming electric
field, but are more concerned with its correlations $\la \eps_j(\vnhat, \nu)
\eps_k^*(\vnhat', \nu') \ra$. The index $j$ runs over the polarisations of the
incoming electric field, which is described in terms of an orthogonal basis on
the sphere. In this work, we will use the conventional decomposition along a basis
in $\hat{\vec{\phi}}$ and $\hat{\vec{\theta}}$.
In most cases we can treat the emission as
incoherent and originating in the far field such that it is described by an
intensity matrix
\begin{equation}
    \la \eps_j(\vnhat, \nu) \eps_k^* (\vnhat', \nu') \ra =
    \delta^{\refchange{2}}(\vnhat - \vnhat') \delta(\nu - \nu')
    \frac{k_B\nu^2}{c^2} I_{jk}(\vnhat, \nu)\;,
\end{equation}
which we express as a brightness temperature. \refchange{Note that since
$\int\delta^{2}(\vnhat) \dhn = 1$, the units of $\delta^{2}(\vnhat)$ are
inverse steradians.} The above equation can be decomposed in terms of
Stokes parameters $I$, $Q$, $U$ and $V$,
giving
\begin{align}
    I_{jk}(\vnhat) &= \calP_{jk}^I I(\vnhat) + \calP_{jk}^Q Q(\vnhat)
    + \calP_{jk}^U U(\vnhat) + \calP_{jk}^V V(\vnhat)\; ,
\label{eq:EEpol}
\end{align}
where the polarisation matrices $\mathcal{P}^P$ are equal to the Pauli matrices in an orthonormal basis
\begin{align}
\calP^I_{jk} & = \begin{pmatrix} 1 & 0 \\ 0 & 1\end{pmatrix},
&
\calP^Q_{jk} &= \begin{pmatrix} 1 & 0 \\ 0 & -1\end{pmatrix},
\notag \\
\calP^U_{jk} &= \begin{pmatrix} 0 & 1 \\ 1 & 0\end{pmatrix},
&
\calP^V_{jk} &= \begin{pmatrix} 0 & -i \\ i & 0\end{pmatrix}.
\end{align}
For notational convenience we will rewrite \eqref{eq:EEpol} as
\begin{equation}
    I_{jk}(\vnhat) = \calP_{jk}^P I_P(\vnhat)
    \;,
\end{equation}
where there is an implied summation over the polarisation index $P$, which we
use throughout for repeated indices with one raised and one lowered.
We thus have
\begin{equation}
    \label{eq:EEpol2}
  \la \eps_j(\vnhat, \nu) \eps_k^* (\vnhat', \nu') \ra =
  \delta^{\refchange{2}}(\vnhat - \vnhat') \delta(\nu - \nu')
  \frac{k_B\nu^2}{c^2}
    \calP_{jk}^P I_P(\vnhat)
    \;.
\end{equation}

The strength of a single unresolved source, at location $\n_s$, is parametrized
by its \refchange{spectral flux density $F_\nu$}, the power per unit collecting area per unit frequency
(usually quoted in Janskys). This is related to the intensity $I$ as follows.
First, a short E \& M calculation gives the flux per observed solid angle due to
intensity vector $I_P$:
\begin{align}
    \frac{dF_\nu}{d\Omega}
    &= \frac{k_B \nu^2}{c^2} \delta^{jk} \calP^P_{jk} I^s_P(\vnhat,\nu)
    \nonumber\\
    &= \frac{2 k_B \nu^2}{c^2} I^s(\vnhat,\nu)\;.
\end{align}
From this we can read off the unpolarized intensity associated with a single
source \refchange{(which we indicate with a superscript $s$)}:
\begin{align}
I^s(\n,\nu)
  &= \delta^2(\vnhat-\vnhat_s)\frac{c^2}{2 k_B \nu^2} F_\nu \nn \\
  &= \delta^2(\vnhat-\vnhat_s)(3.26 \times 10^{-5} \mbox{\,K})
    \left( \frac{\nu}{\mbox{1\,GHz}} \right)^{-2}
    \left( \frac{F_\nu}{\mbox{1\,Jy}} \right)
\label{eq:Jy_to_K}
\end{align}

The way we have distributed the factors of 2 is such that
for an unpolarized signal the brightness temperature in a single polarization,
say $I_{xx}$, has
the same value as the unpolarized intensity $I$. However, the
flux density in a single polarization has half the value as the total flux.
That is, the unpolarized brightness is the average of the brightnesses in the
individual polarization components, whereas the flux is the sum of the flux of
the polarization components.

\subsection{Antennas and Visibilities}

The signal at an antenna, (normally measured as a digitized voltage)
can be written in terms of these plane
waves and an antenna response function $A_j^a(\vnhat, \nu)$ given by
\begin{equation}
\label{eq:feed}
    \eta_a = \int A_a^j(\vnhat, \nu) \eps_j(\vnhat, \nu) e^{i2\pi
    \vu_a\cdot\vnhat} \dhn + n_a(\nu) \; ,
\end{equation}
where $\vu_a = \vx_a / \lambda$, the feed position given in wavelengths and
$n_i(\nu)$ is the receiver noise. The
antenna response, $A_a^i(\vnhat, \nu)$ is a complex
two-dimensional vector field giving the response to waves of both polarisation
at every location on the sky. The response is normalized such
that
\begin{equation}
    \label{e:pri_beam_norm}
    \int \delta_{jk} A_a^j(\vnhat, \nu)A_a^{k}(\vnhat, \nu)^* d^2\vnhat = 1\;.
\end{equation}
\refchange{Note that in other works the response is often normalized such that its maximum
value is unity.}
\refchange{The quantity $D(\vnhat) \equiv \delta_{jk} A_a^j A_a^{k*}$ is the
directivity \citep{6758443}, which is related to the effective area of the
antenna by}\footnote{
\refchange{We will assume a calibration relative to a sky source, and as
such ignore losses parameterized by the radiation efficiency that would
normally enter this equation. These losses instead get absorbed into the
definitions of the noise properties (\textsl{i.e.} $T_r$, defined below).
}}
\refchange{$A_{\rm eff} = \lambda^2 D$.}
For a well designed antenna, the effective area is related to the physical area of
the antenna by an efficiency factor of order unity. The effective solid angle
over which the antenna has response---or the beam solid angle---is thus
$\Omega_A \sim \lambda^2/A_{\rm eff}$.

Antennas are often deployed in pairs with complementary polarization response.
That is, for each antenna, there is a second co-located antenna with a response
that has a similar angular and frequency dependence but nearly orthogonal
dependence in polarization space (index $i$). We treat these as distinct
antennas with different $a$ indices.

The quantity recorded by most radio interferometers is the visibility, the
correlation between a pair of feeds. This is evaluated by estimating the
covariance between feeds over a set of time samples
\begin{align}
    \label{e:Vab}
    V_{ab} &\equiv
    \frac{c^2}{k_B \nu^2}
    \frac{1}{n_\text{samp}}\sum_t {\eta_a[t] \eta_b[t]^*}\\
    C_{ab} &\equiv \la V_{ab} \ra = S_{ab} + N_{ab}\; ,
\end{align}
where $\eta_a[t]$ are discrete time samples of the antenna signals $\eta_a$, and the
total number of samples we are averaging is $n_\text{samp} \equiv \Delta t
\Delta \nu$ samples in time. In the second line we have separated the expected
visibility into contributions from the sky $S_{ab}$, and receiver noise $N_{ab} =
\la n_a n_b^* \ra$. Combining with \eqref{eq:feed} and \eqref{eq:EEpol2} we can
express the measured visibility as
\begin{equation}
    S_{ab}(\nu) = \int A_a^j(\vnhat, \nu) A_b^k(\vnhat, \nu)^* \calP_{jk}^P
    I_P(\vnhat, \nu) e^{i2\pi \vu_{ab} \cdot\vnhat}\dhn\;.
\end{equation}
where $\vu_{ab} = \vu_a - \vu_b$ is the vector separation between the feeds in
wavelengths. With these definitions, if the sky is unpolarized and isotropic
with brightness temperature $T$ (i.e.~$I(\n,\nu) = T$) then the sky
auto-correlation is $S_{aa} = T$.

In the case where the sky contains a single unpolarized point source, 
we can combine the
above with Equation~\ref{eq:Jy_to_K} to obtain
\begin{equation}
    \label{eq:V_point_source}
    S^{s}_{ab}(\nu)
    =
    \frac{c^2}{2 k_B \nu^2} F_\nu
    \delta_{jk}
    A_a^j(\vnhat_s, \nu) A_b^k(\vnhat_s, \nu)^*
    e^{i2\pi \vu_{ab} \cdot\vnhat_s}\;.
\end{equation}
This yields the notion of the antenna forward gain, the maximum response of the antenna
to a point source $S^{s}_{aa} = G_f F_\nu$ with
$G_f = c^2 \delta_{jk}A_a^jA_a^{k*}/ 2 k_B
\nu^2 = A_{\rm eff} / 2 k_B$, which has units K/Jy.

In many cases we will adopt a simple noise model where receiver noise is
constant, uncorrelated from antenna to antenna, and dominates over the sky:
\begin{align}
    \label{e:simple_noise}
    N_{ab}(\nu) &= T_{\rm r}(\nu)\delta_{ab}
    \nonumber \\
    T_{\rm r} &\gg S_{ab}  &&\text{(simple noise)},
\end{align}
where $T_r$ is the receiver noise temperature. However, most results will be
presented in as general form as possible to facilitate extensions.

The covariance of the visibilities is
\citep{1989AJ.....98.1112K, 2015A&C....12..181M}:
\begin{align}
    \label{e:covV}
    \Cov(V_{ab}, V_{cd}) = \frac{C_{ac} C_{bd}^*}{\Delta \nu \Delta t} \;.
\end{align}
This is convenient since it often suffices to use $V_{ab}$ as an estimate of
$C_{ab}$ in the above formula, permitting the covariance to be calculated
directly from the data. Such a scheme is valid even if the visibilities are
uncorrelated.
A special case of the above equation is the variance of an auto-correlation
($a=b=c=d$), where the equation reduces to
\begin{equation}
    \label{e:auto_var}
    \Var(V_{aa}) = \frac{\la V_{aa} \ra^2}{\Delta \nu \Delta t} \;.
\end{equation}
In the case where the receiver noise dominates over the sky and is
described by the simple system temperature model above,
\eqref{e:covV} reduces to the
familiar radiometer equation:
\begin{align}
    \Cov(V_{ab}, V_{cd}) = \delta_{ac}\delta_{bd}
    \frac{T_{\rm r}^2}{\Delta \nu \Delta t}
    &&\text{(simple noise)}.
    \label{e:simple_noise_cov}
\end{align}

\needspace{5ex}

\section{Beamforming}

We will define a \refchange{beamformed visibility}\footnote{%
\refchange{We will henceforth simply use ``beam'' to refer to a beamformed
visibility. This is somewhat inconsistent with standard radio interferometry
where ``beam'' usually refers to the beam response function
(Equation~\ref{e:response_def}); however such a definition becomes
overly verbose in the present work.
}}
as any linear combination of the visibilities:
\begin{equation}
b = w^{ab}V_{ab}\;,
\end{equation}
where we have defined the visibility space beamforming weights $w^{ab}$.
We choose the beams to be normalized such that
\begin{equation}
    \label{e:w_norm}
    \sum_{ab} w^{ab}w^{ab\,*} = 1\;.
\end{equation}
We will see in a moment that in the simple noise model, this normalization
gives the beams the same variance as the visibilities.
\refchange{The beam's expectation value in terms of the sky and noise is}
\begin{equation}
    \la b(\nu) \ra = \int B^{\jindex\kindex}(\vnhat, \nu)
    \calP_{\jindex\kindex}^P
    I_P(\vnhat, \nu) \dhn + w^{ab} N_{ab} \;,
    \label{e:response_def}
\end{equation}
with the \refchange{beam response function (or beam shape function)}
\begin{equation}
    B^{\jindex\kindex}(\vnhat,\nu)
    =
    w^{ab} A_a^i(\vnhat, \nu) A_b^j(\vnhat, \nu)^*
    e^{i2\pi \vu_{ab} \cdot\vnhat}\;.
\end{equation}

The covariance of two formed beams is
\begin{align}
    \mathcal{C}_{XY}
    &\equiv
    \Cov(b_X, b_Y)
    \nonumber
    \\
    &=
    \frac{w^{ab}_X C_{ac} C_{bd}^* w^{cd*}_Y}{\Delta \nu \Delta t} \;,
\label{e:beam_cov}
\end{align}
which in the case of the simple noise model is
\begin{align}
    \mathcal{C}_{XY} = \frac{T_{r}^2 \sum_{ab} w^{ab}_X  w^{ab*}_Y}{\Delta \nu \Delta t}
    \qquad\text{(simple noise)}.
\label{e:beam_cov_simple}
\end{align}

A special class of beams can be formed pre-correlation on the antenna signals,
which can then be squared and integrated. These are termed factorizable beams,
since
their defining feature is that in visibility space the
beam weights factorize:
\begin{align}
\label{e:bf}
b_f
    &=
    \frac{c^2}{k_B \nu^2}
    \frac{1}{n_\text{samp}}\sum_t{|w_f^{a} \eta_{a}[t]|^2}
    \nonumber \\
    &=
    \frac{c^2}{k_B \nu^2}
    \frac{1}{n_\text{samp}}\sum_t{w_f^{a} \eta_{a}[t] w^{b*}\eta_{b}[t]^*}\;.
\end{align}
Hence for factorizable beams we have
\begin{equation}
    \label{e:wf}
    w_f^{ab} = w_f^{a} w_f^{b*}\;.
\end{equation}
Since factorizable beams are the magnitude square of a linear combination
of the pre-correlation antenna signals, they are strictly positive and have
no negative lobes. In analogy to \eqref{e:auto_var}, factorizable beams have the
property that
\begin{equation}
    \label{e:var_bV}
    \Var(b_f) = \frac{\la b_f \ra^2}{\Delta \nu \Delta t} \;.
\end{equation}

Note that the individual antenna patterns $A_a^i$ have been normalized such that
their intensity response integrates over angles to unity
(Equation~\ref{e:pri_beam_norm}).
There is no such relation for formed beams, where $\delta_{jk} B^{jk}$ may
integrate to a
quantity either less than or greater than unity. Our formalism is general
enough to, for instance,
describe beams
that are the difference of two redundant visibilities, which would have no sky
response.
One case where the sky response
does integrate to unity is for factorizable beams when the antenna
\refchange{responses}
are isotropic.

\refchangetwo{Throughout this article several classes of beams are discussed,
using different choices of the weights to achieve different goals. To orient
the reader, we summarize these in Table~\ref{t:beams}.
}

\begin{table*}
\refchangetwo{%
\begin{center}
\begin{tabular}{l l c p{0.5\linewidth}}
    Name & Symbol & Section Ref. & Description \\[\valsep]
    \hline
    Factorizable beam & $b_f$ & \ref{s:pointed} &
        Class of beams that can be formed pre-correlation on antenna signals.
        \\[\valsep]
    Pointed beam & $b_p(\vnhat_p)$ & \ref{s:pointed} &
        Maximum response to direction $\vnhat_p$. Factorizable for identical
        antennas.
        \\[\valsep]
    Optimal pointed beam & $b_\textrm{opt}(\vnhat_s)$ & \ref{s:pointed} &
        Maximum signal-to-noise ratio to source at sky location $\vnhat_s$, same
        as pointed beams for the simple noise model.
        \\[\valsep]
    FFT beams & $b_A$ & \ref{s:fft_beamforming} &
        Efficiently formed set of beams for redundant arrays. Factorizable and
        equivalent to pointed beams at fixed pointing angles $\theta_A$.
        \\[\valsep]
    Naive windowed beam & $b_\textrm{nw}$ & \ref{s:windowed_beams} &
        Antenna space tapered aperture for suppressing sidelobes. Factorizable.
        \\[\valsep]
    Optimal windowed beam & $b_\textrm{ow}$ & \ref{s:windowed_beams} &
        Identical beam shape to naive windowed beam, but higher response. Not
        factorizable.
        \\[\valsep]
\hline
\end{tabular}
\end{center}
\caption{\label{t:beams} Summary of classes of beams discussed.}
}
\end{table*}

\subsection{Pointed Beams}
\label{s:pointed}

When studying discrete, unresolved, unpolarized sources on the sky, one often
wants to maximize
response of the array to a single point at steering angle $\vnhat_p$.
Such a beam signal weights the visibilities (based on
\eqref{eq:V_point_source}, setting $\vnhat_s$ to $\vnhat_p$) and adds
them in phase. This yields the definition of a pointed beam:
\begin{align}
b_p(\vnhat_p) &= \frac{1}{\cal{N}}\sum_{ab} V_{ab}
    \delta_{\jindex\kindex} A_a^\jindex(\vnhat_p, \nu)^* A_b^\kindex(\vnhat_p, \nu)
    e^{-i2\pi \vu_{ab} \cdot\vnhat_p}
    \;,
\end{align}
where,
\begin{align}
\mathcal{N}^2 &\equiv \sum_{ab} \delta_{\jindex\kindex} A_a^i(\vnhat_p, \nu)^* A_b^j(\vnhat_p, \nu)
           \delta_{lm} A_a^l(\vnhat_p, \nu) A_b^m(\vnhat_p, \nu)^*\;.
\end{align}
The corresponding beam weights are thus
\begin{align}
w_p^{ab}(\vnhat_p)
    &= \frac{1}{\cal{N}}\delta_{\jindex\kindex} 
    A_\aindex^\jindex(\vnhat_p, \nu)^* A_\bindex^\kindex(\vnhat_p, \nu)
    e^{-i2\pi \vu_{ab} \cdot\vnhat_p}\;.
    \label{e:wp}
\end{align}
Note that in this general case the weights cannot be factorized
and such a beam cannot be formed
from pre-correlation voltages. This is because a general array can have polarization
response that varies from antenna to antenna and, after contracting with
$\delta_{\jindex\kindex}$, $w_p^{ab}$ will be rank two (the sum of
two factorizable sets of weights).
That is to say polarization information must be summed post correlation.

The gain of the pointed beam is $G_p(\vnhat_p) = c^2 \mathcal{N} / 2 k_B \nu^2$. In the
special case where all antenna beams are identical and the source is at
boresight, then this is just $\nant G_f$, the number of antennas times the
single antenna forward gain.

Pointed beams, as defined here, maximize the signal from a particular point on the
sky, however, they are not optimal in that they do not necessarily maximize the
signal to noise ratio (except in the simple noise model in
\eqref{e:simple_noise}). For factorizable beams, we have Equation~\ref{e:var_bV},
saying that the noise in a beam is proportional to its total power, including
sky and receiver noise contributions.
For non-trivial sky and receiver noise, there may be
sensitivity gains from tuning the beams to remove other signals (receiver cross
talk, Galactic emission, etc.) in favour of
the source of interest. For instance, it may be beneficial for the beam to null
the location of an extraneous bright source to prevent that source from adding noise.
To find the optimal beam we write the
signal as $\langle b^s \rangle = w^{ab} S^{s}_{ab}$ and the noise as
$(\Delta b)^2 =
w^{ab} C_{ac} C^*_{bc} w^{cd}/(\Delta \nu \Delta t)$
(Equation~\ref{e:beam_cov}) and maximize $(\langle b^s \rangle/\Delta b)^2$
with respect
to $w^{ab}$ by setting the derivatives to zero. This yields
\begin{equation}
    S^s_{ab} - \frac{\langle b_\textrm{opt}^s\rangle}{(\Delta b_\textrm{opt})^2 \Delta \nu \Delta t}
    C_{ac} C^*_{bd} w_{\rm opt}^{cd*} = 0
    \;.
\end{equation}
The prefactors of the second term have no dependence on the antenna index $a$
and are therefore irrelevant. Thus we have
\begin{equation}
    \label{e:wopt}
    w^{ab}_{\rm opt} \propto \sum_{cd}C^{-1*}_{ac} S_{cd}^{s*} C^{-1}_{bd}
    \;.
\end{equation}
\refchange{Note that in the simple noise model, $C_{\aindex\bindex}$ is
diagonal, and Equation~\ref{e:wopt} reduces to the weights for the pointed beam
in Equation~\ref{e:wp}.}
Such optimal beams are mathematically cumbersome, and as such, we will work mostly
with pointed beams. The exception is in Section~\ref{s:complicated_noise}.

\subsection{Redundant arrays and Fourier transform beamforming}
\label{s:fft_beamforming}

We will now restrict the discussion to redundant arrays of antenna. These are
arrays with identical antenna patterns and regular spacings. For simplicity we
will consider identical, single-polarization antennas such that
$A_a^i(\vnhat, \nu)=A^i(\vnhat,\nu)$ is independent of $a$, a linear (1D)
array such that $\vu_{ab} = \vxhat d (a - b)/\lambda$, and observing a 1D sky
such that $\vnhat \cdot \vxhat = \sin(\theta)$ where $\theta$ is the 1D zenith
angle. As such, the sky
contribution to the visibilities $S_{ab}$ depends only on the antenna
separation $(a - b)$.
We also assume that the noise $N_{ab}$ depends only on $(a - b)$, the
simple noise model in Equation~\ref{e:simple_noise} being a special case.

\begin{figure*}
    \begin{center}
        \includegraphics[scale=0.7]{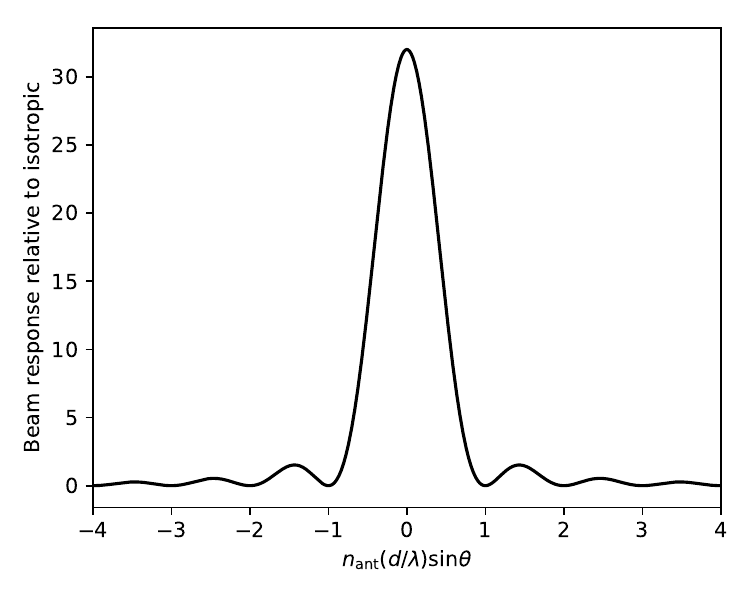}
        \includegraphics[scale=0.7]{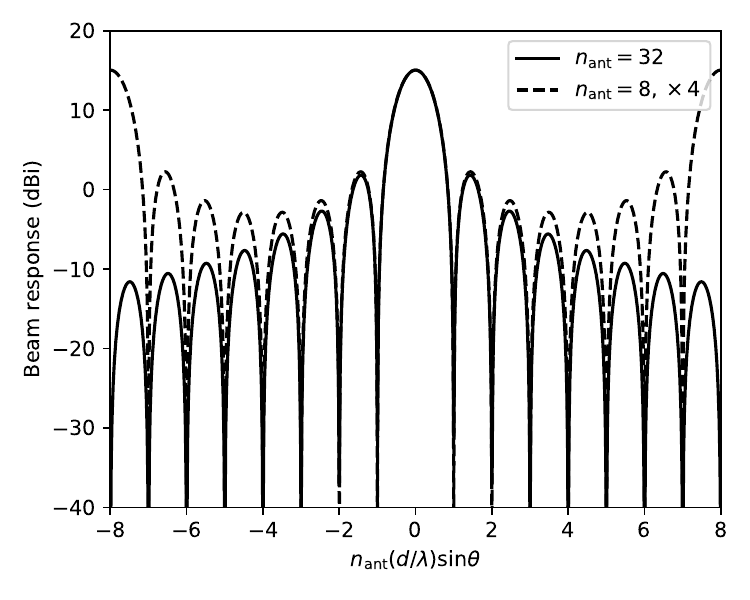}
        \caption{\label{f:beams} The response function of the beamformed
        visibilities for a
        regular linear array with 32 elements, on linear (left) and logarithmic
        (right) scales, neglecting the primary antenna response (assuming
        $A^i(\vnhat)$ is isotropic).
        The horizontal axis is scaled to be in units of the natural beam
        width $\lambda/(\nant d)$. \refchange{To illustrate the effects of aliasing, in
        the right panel we also show the response for an array with 8 element,
        scaled by a factor of 4 to match the peak response of the 32-element
        array.}
        }
    \end{center}
\end{figure*}

With these simplifications, the pointed beam, which for the simple noise model
is also the optimal beam, becomes
\begin{equation}
    b_p(\theta_p) = \frac{1}{\nant}\sum_{ab}e^{-i2\pi(a-b)
(d/\lambda)\sin\theta_p}V_{ab}\;.
\label{e:bp}
\end{equation}
and thus
\begin{equation}
w_p^{ab}(\theta_p) = \frac{1}{\nant}e^{-i2\pi
    (a - b)(d/\lambda)\sin \theta_p}\;.
\end{equation}
The weights are factorizable such that
\begin{equation}
w_p^{\aindex}(\theta_p) = \frac{1}{\sqrt{\nant}}e^{-i2\pi
    \aindex(d/\lambda)\sin\theta_p}\;.
\label{e:c_pointed}
\end{equation}
\refchange{This form can be understood from Equation~\ref{e:wp} by noticing
that the weights no longer need to depend on $A_a^i$ (which are now independent
of $a$), and recomputing the normalization.}
The \refchange{response function} of such a beam is
\begin{align}
    \label{e:pointed_beam_shape}
    B^{\jindex\kindex}_p(\theta)
    &=
    A^\jindex(\theta) A^\kindex(\theta)^*
    \frac{1}{\nant} \sum_{\aindex\bindex}e^{-i2\pi (\aindex -
    \bindex)(d/\lambda)(\sin\theta_p -
    \sin\theta)}
    \nonumber\\
    &=
    \frac{A^\jindex(\theta) A^\kindex(\theta)^*}{\nant}
    \frac{\sin^2\left[\nant\pi(d/\lambda)(\sin\theta_p - \sin\theta)\right]}{%
        \sin^2\left[\pi(d/\lambda)(\sin\theta_p - \sin\theta)\right]}\;.
\end{align}
For angles close to the pointing
angle compared to the alias limit
(that is for $(d/\lambda) (\sin\theta_p - \sin\theta) \ll 1)$), or equivalently
the limit of closely spaced antennas,
the function that multiplies the antenna patterns approximates the familiar
sinc-squared function expected for a square aperture:
\begin{align}
    \label{e:sinc_approx}
    &\frac{\sin^2\left[\pi\nant(d/\lambda)(\sin\theta_p - \sin\theta)\right]}{%
        \sin^2\left[\pi(d/\lambda)(\sin\theta_p - \sin\theta)\right]}
        \nonumber\\
        &\qquad\approx
        \nant^2 \sinc^2 \left[\pi\nant(d/\lambda)(\sin\theta_p - \sin\theta) \right]\;.
\end{align}
Note that unlike a continuous square aperture, the beam in
Equation~\ref{e:pointed_beam_shape} has aliases---additional directions of high
response---for $\sin\theta_s - \sin\theta$ equal to multiples of $\lambda/d$.
We show this beam shape in Figure~\ref{f:beams}.

While in the simple noise model the error in the visibilities is uncorrelated,
the error in pointed
beams is not, and
\begin{align}
    \label{e:pointed_beam_covar}
    &\Cov( b_p(\theta_p) b_p({\theta_p}'))
    \nonumber\\
    &=
    \frac{T_{r}^2}{\Delta \nu \Delta t}
    \frac{1}{\nant^2}
    \sum_{\aindex\bindex}
    e^{-i2\pi (\aindex - \bindex)(d/\lambda)(\sin\theta_p - \sin{\theta_p}')}
    \nonumber\\
    &=
    \frac{T_{r}^2}{\Delta \nu \Delta t}
    \frac{1}{\nant^2}
    \frac{\sin^2\left[\nant\pi(d/\lambda)(\sin\theta_p - \sin{\theta_p}')\right]}{%
        \sin^2\left[\pi(d/\lambda)(\sin\theta_p - \sin{\theta_p}')\right]}
    \nonumber\\
    &\text{(simple noise)}
\end{align}

This has a similar functional form to the beam shape
(Equation~\ref{e:pointed_beam_shape}). It is zero if
$(\sin\theta_p - \sin{\theta_p'})$ is a multiple of
$\lambda/(\nant d)$ and
non-trivial otherwise.

Equation~\ref{e:c_pointed} hints that many beams could be efficiently formed
using a spatial FFT of the pre-correlation antenna signals, $\eta_a[t]$.
$M$ beams can be formed by zero padding the $\nant$ antenna
to length $M$ and taking an FFT in the spatial direction \refchange{(over index
$a$)}. If $M < \nant$ beams
are desired then rather than zero padding, the array should be populated by
cyclically co-adding the
signals from the antennas. That is, the $A^{\textrm{th}}$ element of the array
to be Fourier transformed should be the sum of all the $\eta_a$ with
$a\pmod M = A$. We then have:
\refchange{
\begin{equation}
    \label{e:fftbeamform}
    b_p(\theta_A) = \frac{1}{n_\text{samp}}\sum_t \left| \sum_a \eta_a[t] \frac{1}{\sqrt{\nant}}e^{-i2\pi
    A\aindex/M} \right|^2\;,
\end{equation}
}
noting that the inner sum over $a$ is an FFT. The equivalent voltage
beamforming weights are
\begin{equation}
    \label{e:w_A^a}
    w_p^\aindex(\theta_A) = \frac{1}{\sqrt{\nant}}e^{-i2\pi A\aindex/M}\;.
\end{equation}
and the discrete steering angles of the formed beams are
\begin{equation}
    \label{e:theta_A}
    \sin \theta_A = \frac{A \lambda}{Md}\;.
\end{equation}
This equation is valid for any $A$ satisfying the constraint
$|\sin \theta_A| \le 1$, with those outside the $0$ to $M - 1$ range describing
aliases of the $A \pmod M$ beams.
\refchange{%
We will thus refer to the $b_p(\theta_A)$ (hereafter simply $b_A$) as the
Fourier trasform beams or FFT beams.
Note in the above equation that the locations of the beams are wavelength dependent, so,
without modification, the FFT beams are not appropriate fan beams
for searches for broadband point sources such as FRBs.}
If $M = \nant$ then the beams have independent errors for the simple noise
model
(Equation~\ref{e:pointed_beam_covar}), but this is not
the case in general.

\subsection{Redundancy-stacked visibilities}

Equation~\ref{e:bp} can be rewritten as
\begin{equation}
\label{e:b_V_delta}
b_p(\theta_p) = \frac{1}{\nant}\sum_{\delta=-(\nant-1)}^{\nant-1}e^{-i2\pi
\delta (d/\lambda)\sin\theta_p}
\widetilde{V}_\delta,
\end{equation}
where
\begin{equation}
    \widetilde{V}_\delta
    \equiv \sum_{\alphaindex=\max(0, -\delta)}^{\min(n_{\rm ant}, n_{\rm ant} - \delta)}
    V_{\alphaindex+\delta\,\alphaindex}.
\end{equation}
Here, $\delta$ indexes the difference between two feed indices $\aindex -
\bindex$, and the $\alphaindex$ index runs over the redundant pairs (we use
$\alphaindex$ over $\aindex$ to make it clear that the index limits are
different and dependant on $\delta$).
The quantity $\widetilde{V}_\delta$ is the sum of the $\nant-|\delta|$ visibilities
whose baselines are redundant.
Note that because the sky contribution to the
visibilities is the
same for redundant baselines \refchange{($V_{ab}$ depends only on $a - b$)},
$\widetilde{V}_\delta$ contains all the information in a redundant array
in the case of the simple noise model where the visibilities are
uncorrelated.
This is not the case for non-trivial noise or
non-negligable contributions to the visibility uncertainty from the sky, where the
visibilities are correlated (Equation~\ref{e:covV}) and that correlation is
visibility dependant even amongst redundant pairs.  That is, the correlation
between $V_{\aindex=2,\bindex=1}$ and $V_{\aindex=3,\bindex=2}$
will not be the same as that between $V_{\aindex=2,\bindex=1}$ and
$V_{\aindex=4,\bindex=3}$, and an optimal sum of these three visibilities
must take into account these correlations.
\refchange{
To get an idea of how severe the information loss could be, we have considered
toy models where visibilities are dominated by a
single sky structure resolved by roughly half the baselines. We find, that
the increase in uncertainty on the stacked visibilities
can be of order unity compared
to an optimally weighted stack. However, the information loss
remains to be quantified for a realistic sky and instrument.
}

Nonetheless, these correlations are small in most systems where the
auto-correlations ($\aindex = \bindex$) are much larger in amplitude than the
cross-correlations ($\aindex \neq \bindex$). We
will thus assume that $\widetilde{V}_\delta$ contains essentially all the
information from the array hereafter. As such most beams of interest can be formed
directly in this space.
%
%
We will denote the weights in such cases as\footnote{%
\refchange{The symbols for beamforming weights in voltage space ($w^a$),
redundancy-stacked visibility space ($w^\deltaindex$), and the later defined
FFT beam space ($w^A$) are distinguishable only by the type of character used as an
index. This notation is convenient but care must be taken to not confuse the
weights in different spaces.
}}
$w^\deltaindex$, such that $b = w^\deltaindex \widetilde{V}_\deltaindex$.
\refchange{
These
are trivially related to beamforming weights in unstacked visibility space:
\begin{align}
    w^\deltaindex &= w^{\aindex=\alphaindex+\delta,\bindex=\alphaindex}
    \;.
\end{align}
}

For the Fourier transform formed beams, Equation~\ref{e:b_V_delta} becomes
\begin{align}
    b_A &= \frac{1}{\nant}\sum_{\aindex\bindex}e^{-i2\pi
    A(\aindex-\bindex)/M}
    V_{\aindex\bindex}
    \nonumber
    \\
    &= \frac{1}{\nant}
    \sumdelta
    e^{-i2\pi A\delta/M}
    \widetilde{V}_\delta
    \;.
\label{e:b2n-1}
\end{align}
Equation~\ref{e:b2n-1} indicates that for $M \geq 2\nant - 1$, $b_p(\theta_A)$ and
$\widetilde{V}_\delta$ are related by a discrete Fourier transform (with
$\widetilde{V}_\delta$ zero padded to length $M$). Since Fourier transforms
are invertable,
$b_A$ contains the same information as $\widetilde{V}_\delta$.
That the minimum number of FFT formed
beams for which this is true is $M = 2\nant - 1$ agrees with the number of
degrees of freedom in $\widetilde{V}_\delta$.
Because $\widetilde{V}_\delta = \widetilde{V}_{-\delta}^*$
there are $\nant$ independent, but complex, numbers. Since
$\widetilde{V}_{\delta=0}$ is
real, there are $2\nant-1$ degrees of freedom. The number of independent beams
can also be understood in terms of convolution theorem,
where squaring the spatially transformed uncorrelated input signals
is equivalent to a spatial
auto-convolution of those signals. Padding to $2\nant-1$ is required to
deal with the non-periodicity of the antenna array. This also makes it clear
that padding to any number larger than $2\nant-1$ also preserves
information. This is convenient since $M = 2\nant$ is likely more factorizable
and can thus be implemented more efficiently with a fast Fourier transform.

As such, FFT beamforming provides a method to correlate the antenna signals,
\refchange{since the $b_A$ can be formed using an FFT to implement
Equation~\ref{e:fftbeamform}
which scales as $\nant \log{\nant}$ rather than $\nant^2$}, and this method
contains the same information as the redundancy-stacked visibilities
$\widetilde{V}_\delta$. Both the FFT beamforming and the redundancy stacking
are information preserving in the case where
the visibility auto-correlations are the dominant contributions to the
visibility uncertainty as discussed above.

Since the FFT beams have the same information content as the
$\widetilde{V}_\delta$, it is clear that any beam shape that can be produced
in visibility space can also be
achieved by taking linear combinations of the FFT
beams.
This has a small computational cost
compared to the initial FFT beamforming due to the typically high degree of
$\Delta t\Delta \nu$ downsampling in intensity space \refchange{(here we refer to the sums
over $t$ in Equations~\ref{e:Vab}, \ref{e:bf}, and \ref{e:fftbeamform})}.
Being able to form a beam with any shape is not
equivalent to being able to form any beam. For instance, a
beam with $w^{a=2,b=1} \neq w^{a=3,b=2}$ cannot be formed since $V_{a=2,b=1}$
and $V_{a=3,b=2}$ each
contribute to $\widetilde{V}_{\delta=1}$ with equal weight.
However, such beams are clearly non-optimal since $V_{a=2,b=1}$ and $V_{a=3,b=2}$ contain
the same sky information and independent noise realizations.

Assuming hereafter that $M \geq 2\nant - 1$, the $b_A$ provide an
alternate basis for forming any beam where $w^{\aindex\bindex}$ depends only
only on $\delta = \aindex-\bindex$.
We will denote the coefficients in this space as $w^A$, such
that such a beam can be written
\begin{equation}
    b = \sum_{A=0}^{M-1} w^A b_A
    \;.
\end{equation}
The inverse of Equation~\ref{e:b2n-1} is
\begin{equation}
    \widetilde{V}_\delta = \frac{\nant}{M} \sum_A e^{i2\pi\delta A/M}
    b_A
    \;.
\label{e:inv_b2n-1}
\end{equation}
and from substituting this equation into
$b = w^\deltaindex \widetilde{V}_\deltaindex$
it can be shown that
\begin{equation}
    \label{e:wA_from_wdelta}
    w^A = \frac{\nant}{M} \sum_\delta e^{i2\pi \delta A/M}
    w^\deltaindex\;,
\end{equation}
and likewise
\begin{equation}
    w^\deltaindex = \frac{1}{\nant} \sum_A e^{-i2\pi \delta A/M}
    w^A\;.
\end{equation}

\begin{figure}
    \begin{center}
        \includegraphics[scale=0.65]{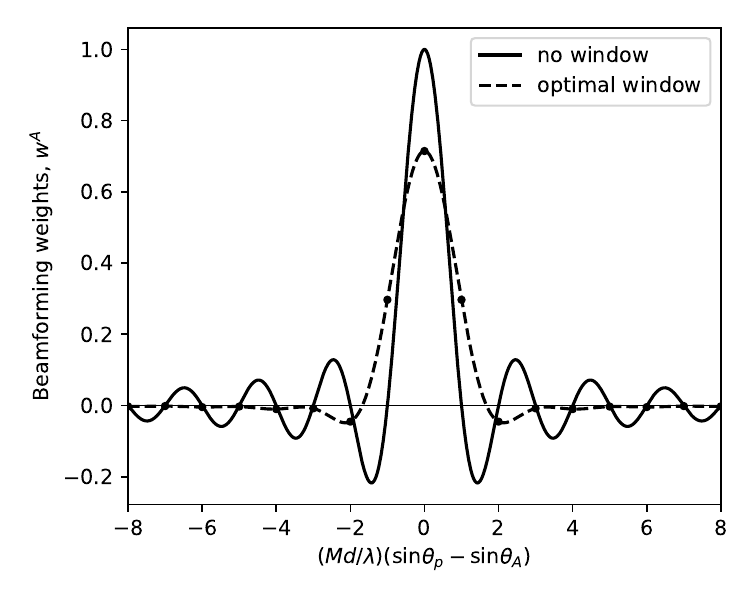}
        \caption{\label{f:wA}
        Weights, $w^A$, for forming a beam pointed to an arbitrary location from FFT
        beams \refchange{(Equation~\ref{e:bp_max_arbitrary})}. The horizontal axis has been scaled to be in units of
        the FFT beam
        steering angle
        separation $\lambda/(Md)$. The windowed version
        \refchange{(Equation~\ref{e:wA_windowed}, to be described later in
        Section~\ref{s:windowed_beams})} uses the
        same half-sine-wave window as
        in Figure~\ref{f:windowed_beams}. Note that the `naive' windowed beam
        cannot be formed in this space. The total number of formed beams has
        been set to $M=2\nant - 1$ with $\nant=32$.
        }
    \end{center}
\end{figure}

As an example of forming an arbitrarily shaped beam from the FFT beams,
a pointed beam to arbitrary steering angle
$\theta_p$ can be
formed.
\refchange{That is, pointed beam locations can be ``regridded'' to angles other
than the FFT steering angles $\theta_A$.}
Combining Equations~\ref{e:b_V_delta} and~\ref{e:inv_b2n-1} we have,
\begin{align}
\label{e:bp_max_arbitrary}
    w^A_p(\theta_p) &= \frac{1}{M}
    \frac{\sin\left[(2\nant -1)\pi y_p^A\right]}{\sin(\pi y_p^A)}
    \nonumber\\
    y_p^A &\equiv (d/\lambda) \sin \theta_p - A/M
    \;.
\end{align}
These weights $w^A_p(\theta_p)$ are beam
regridding coefficients, whose functional form is also approximated by 
a $\sinc$ function \refchange{(Equation~\ref{e:sinc_approx})}. These
coefficients are
shown in Figure~\ref{f:wA}.
\refchange{%
Among other applications, this
solves the location-chromaticity problem for fan beam implementations that use
FFT beamforming, since the FFT beams can be regridded to arbitrary and
achromatic steering angles on a frequency-by-frequency basis.
}

As a final note, in visibility space we have Equation~\ref{e:covV} which
permits the covariance of the visibilities to be estimated from the
visibilities themselves for arbitrary noise and sky. This is also true of the
FFT formed beams where Equation~\ref{e:beam_cov} can be used with $C_{ab}$ written
in terms of $\la b_A \ra$ using Equation~\ref{e:inv_b2n-1}. This however does
not yield a compact expression and is best calculated numerically.

\subsection{\refchange{Non-redundancy}}

\label{s:nonredundancy}
\refchange{%
Prior to delving into applications of our formalism it is important to evaluate the
validity of the assumptions made in this section. Of particular concern
is the assumption of redundancy: that all antennas are equally spaced and have
the same response to the sky. This may be broken due to antenna-to-antenna
variations in the response functions $A^i_a$, from calibration errors in
analogue and digital stages of the signal chains, or from departures from
regularity of the antenna locations. We will analyse the effects
of these variations by considering the sensitivity of a formed beam to a point
source, which is the most relevant measure for time-domain radio astronomy.
For illustrative purposes, we will first consider variations that affect the signal
part of the visibilities but not the noise, such as variations in the antenna
responses or departures of the antenna locations from regular spacings.
Calibration errors in the analogue chains multiply both signal and noise,
which we will consider later.
}

\refchange{%
We model effects of these feed-to-feed variations by mapping the signal
contribution to the visibilities
\begin{equation}
\label{e:vis_pert}
S^s_{\aindex\bindex} \rightarrow
    \tilde{S}^s_{\aindex\bindex} \equiv
    e^{\gamma_\aindex + i\psi_\aindex}
    e^{\gamma_\bindex - i\psi_\bindex}
    S^s_{\aindex\bindex}
\;,
\end{equation}
where the $\gamma_\aindex$'s and $\psi_\aindex$'s are real numbers representing
variations in the point-source response in
amplitude and phase respectively.
These variations are assumed to be pertubatively small and we use the above
exponential parameterization for algebraic convenience.
}

\refchange{%
We now calculate how a source's expected contribution to a pointed beam
($\langle b_p^s \rangle \equiv w_p^{\aindex\bindex} S^s_{\aindex\bindex}$) is
affected by these perturbations to the visibilities
(the perturbed contribution will be denoted by $\langle \tilde{b}_p^s \rangle$).
Setting $V_{\aindex\bindex}$ to $\tilde{S}^s_{\aindex\bindex}$ in
Equation~\ref{e:bp}, expanding to
second order in the response variations,
and substituting Equation~\ref{eq:V_point_source}, we find
\begin{equation}
    \langle\tilde{b}_p^s\rangle \approx \langle b_p^s \rangle
    \left[%
        1 + 2\bar{\gamma} + 2\bar{\gamma}^2 + (\Delta\gamma)^2 - (\Delta\psi)^2
    \right]\;.
\end{equation}
Here $\bar{\gamma} \equiv \sum_\aindex \gamma_\aindex / \nant$ (the mean of
$\gamma_\aindex$ over antennas),
$(\Delta\gamma)^2 \equiv \sum_\aindex (\gamma_\aindex - \bar{\gamma})^2 /
\nant$ (the variance), and likewise for $\psi$.
}

\refchange{%
In the above equation, the terms with $\bar\gamma$ and $\bar{\gamma}^2$
represent departures of the mean response from the nominal value but do not
represent antenna-to-antenna variations and thus have no bearing on the present
discussion on departures from redundancy. The true departures affect the
overall sensitivity to the source at second order, and thus the sensitivity is
rather robust to these variations. For example, 10\% RMS variations in the
amplitude response, or 0.1 radian RMS variations in the phase response affect the
point-source sensitivity by 1\%, a tolerable change in many applications and
for a readily achievable uniformity in antenna response. Specifically, the
CHIME Pathfinder has achieved roughly 10\% antenna response uniformity
\citep{2016SPIE.9906E..0DB}.
Surprisingly,
variations in the amplitude response actually increase the sensitivity to point
sources, albeit by a small amount.
}

\refchange{
Likewise, calibration errors can
be treated in a similar way and, while the effect on signal will be identical
to the above calculation,
the noise will also be affected. Thus we will substitute
\begin{equation}
\label{e:noise_pert}
N_{\aindex\bindex} \rightarrow
    \tilde{N}_{\aindex\bindex} \equiv
    e^{\gamma_\aindex + i\psi_\aindex}
    e^{\gamma_\bindex - i\psi_\bindex}
    N_{\aindex\bindex}
\;.
\end{equation}
As before, we define
$\langle b_p^n \rangle \equiv w_p^{\aindex\bindex} N_{\aindex\bindex}$, and the
perturbed version $\langle \tilde{b}_p^n \rangle$. Again we start with
Equation~\ref{e:bp}, this time setting $V_{\aindex\bindex}$ to
$\tilde{N}_{\aindex\bindex}$ and employing the simple noise model in
Equation~\ref{e:simple_noise}. We find
\begin{equation}
     \langle\tilde{b}_p^n\rangle \approx \langle b_p^n \rangle
    \left[%
        1 + 2\bar{\gamma} + 2\bar{\gamma}^2 + 2(\Delta\gamma)^2
    \right]\;.
\end{equation}
For the simple noise model where $\langle b_p^n \rangle \gg \langle b_p^s \rangle$,
and as a consequence of Equation~\ref{e:var_bV},
the signal to noise ratio is
\begin{align}
    \widetilde{\textrm{SNR}}
    &=
    \frac{1}{\sqrt{\Delta\nu\Delta t}}
    \frac{\langle\tilde{b}_p^s\rangle}{ \langle\tilde{b}_p^n\rangle}
    \nonumber\\
    &\approx
    \frac{1}{\sqrt{\Delta\nu\Delta t}}
    \frac{\langle{b}_p^s\rangle}{ \langle{b}_p^n\rangle}
    \left[%
        1 - (\Delta\gamma)^2 - (\Delta\psi)^2
    \right]
    \nonumber\\
    &\approx
    {\textrm{SNR}}
    \left[%
        1 - (\Delta\gamma)^2 - (\Delta\psi)^2
    \right]
    \;.
\end{align}
As such, for calibration errors the loss in point source sensitivity is also
second order in the antenna-to-antenna calibration variations.
}

\refchange{%
While point source sensitivity is not the only relevant metric, we expect other
effects, such as beam shape perturbations, to be of the same order.
As such, the utility of the above formalism and of the
applications presented below is quite robust.
}

\section{Applications}
\label{s:applications}

In this section we apply the formalism developed above to derive several
observationally useful algorithms and to analyse the information content of the
FFT formed beams in several contexts.

\subsection{Controlling beam shape with windowing}
\label{s:windowed_beams}

In some applications it is desirable to control the shape of formed beams to
suppress the large sidelobes apparent in Figure~\ref{f:beams}.
The naive way to
do this is to form a factorizable beam ($b_\textrm{nw}$) that windows
the spatial Fourier transform such that
$w_\textrm{nw}^\aindex \sim h^\aindex w_p^\aindex(\theta_p)$
where $h^a$ is a window function. This effectively tapers the
illumination of the aperture (sacrificing aperture efficiency),
in direct analogue to \refchange{how the illumination of the dish by the
feed affects beam shape in} telescopes that use
optical focussing \cite[Chapter~15.1.2]{2017isra.book.....T}.
The resulting beam shape is
\begin{align}
    \label{e:windowed_beam_shape}
    B_\textrm{nw}^{\jindex\kindex}(\theta)
    &=
    A^\jindex(\theta) A^\kindex(\theta)^*
    \frac{1}{\mathcal{N}\nant}
    \nonumber\\
    &\qquad \times
    \sum_{\aindex\bindex}
    h^\aindex h^{\bindex *}
    e^{-i2\pi (\aindex - \bindex)(d/\lambda)(\sin\theta_p - \sin\theta)}
    \nonumber\\
    &=
    A^\jindex(\theta) A^\kindex(\theta)^*
    \frac{1}{\mathcal{N}\nant}
    \nonumber\\
    &\qquad \times
    \left|
    \sum_{\aindex}
    h^\aindex
    e^{-i2\pi \aindex(d/\lambda)(\sin\theta_p - \sin\theta)}
    \right|^2
    \nonumber\\
    \mathcal{N}^2
    &\equiv \frac{1}{\nant^2}\sum_{\aindex\bindex}
    h^a h^{a*} h^b h^{b*} = \left(\frac{1}{\nant}\sum_{\aindex}
    h^a h^{a*} \right)^2\;.
\end{align}
Comparing to the windowed pointed beams, we see that the windowed beam
shape simply replaces the sinc-squared-like factor in
Equation~\ref{e:pointed_beam_shape} with the square of the Fourier transform
of the window \refchange{(factor in $|\phantom{A}|^2$)}.

Windowing the array in this way is, however, a sub-optimal way to achieve a
given beam shape, since it assigns different weights to redundant visibilities,
which are independent measurements of identical sky information. Such beams cannot
be formed from $\widetilde{V}_\delta$ or $b_A$.

To form a beam with the same shape as that above, but maintaining the maximum amount
of sky signal (the ``optimally'' windowed beam, $b_\textrm{ow}$), we must find weights that
depend only on baseline \refchange{length (ie., depend only on $\delta = a - b$)}, whose sum
over redundant baselines is proportional the same sum for the above weights.
That is:
\begin{align}
    \sumalpha {w_\textrm{ow}^\deltaindex}
    &=
    (\nant - |\deltaindex|) {w_\textrm{ow}^\deltaindex}
    =
    \frac{1}{\mathcal{N}}
    \sumalpha
    h^{\alphaindex+\delta} h^{\alphaindex *}
    w_p^{\deltaindex}(\theta_p)
    \nonumber\\
    w_\textrm{ow}^\deltaindex &=
    \frac{
    w_p^{\deltaindex}(\theta_p)
    }{\mathcal{N}(\nant - |\deltaindex|)}
    \sumalpha
    h^{\alphaindex+\delta} h^{\alphaindex *}
    \;,
    \label{e:w_opt}
\end{align}
where the normalization does not have a simple form but is straight forward to
calculate numerically for a given window and number of elements. \refchange{Notice that
the factor involving a sum over $\alpha$ is the auto-convolution of the window
function.}

\begin{figure*}
    \begin{center}
        \includegraphics[scale=0.7]{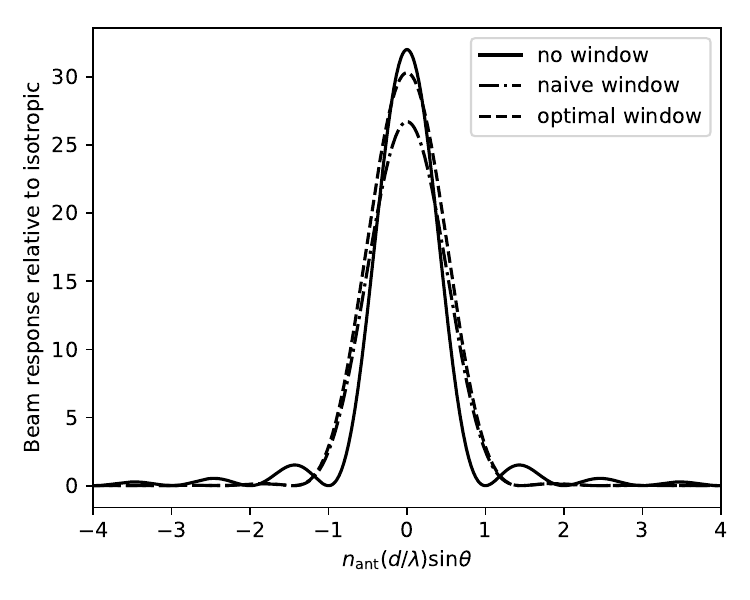}
        \includegraphics[scale=0.7]{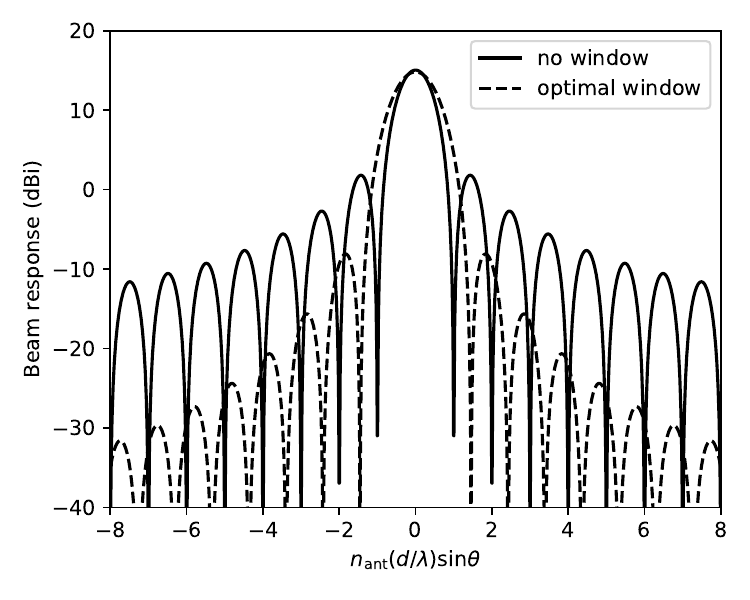}
        \caption{\label{f:windowed_beams} \refchange{The response function of formed beams for a
        regular linear array with 32 elements, on linear (left) and logarithmic
        (right) scales, for different windowing schemes. As in
        Figure~\ref{f:beams}, we neglect the primary antenna response (assuming
        $A^i(\vnhat)$ is isotropic).
        The horizontal axis is scaled to be in units of the natural beam
        width $\lambda/(\nant d)$.
        The beam with no windowing
        maximizes the sky response in the steering direction
        (Equation~\ref{e:pointed_beam_shape}) and is the
        identical curve as in Figure~\ref{f:beams}. Naive windowing
        multiplies the array by a simple half-sine window function prior to
        voltage-space FFT
        beamforming to taper the
        aperture and control side lobes (Equation~\ref{e:windowed_beam_shape}).
        Optimal windowing takes the
        combination of visibilities (Equation~\ref{e:w_opt})
        that achieves the same angular response as naive
        windowing but maximally preserves sky response.
        Since the naive-window and optimal-window curves are
        proportional, the former is omitted from the logarithmic
        plot.}}
    \end{center}
\end{figure*}

As an illustrative example, we use
a simple half-sine wave window function given by
$h^a = \sin [\pi (a + \onehalf) / \nant]$. This particular window function is
relatively broad compared to the commonly used Hann and Blackman
functions, preserving more area and thus more sky information. This is at the
expense of a less gradual taper and thus inferior sidelobe suppression. The
resulting sky response for both the naive window and the optimal window are shown in
Figure~\ref{f:windowed_beams}. For the sine window used here,
the naive windowing has 83\% the peak sky response of the unwindowed pointed beam,
while the optimal windowed beam achieves 95\%.
The optimal window effectively pays a smaller aperture
efficiency price for sidelobe suppression.
\refchange{
Also, while the unwindowed and
naive-windowed beam shapes have the same sky-area integrated response, the optimal
windowed beam has 14\% more, making it \emph{more} sensitive to resolved extended
sources and point-source searches where the flux distribution is shallower than
$N \sim (S_\nu^{\rm min})^{-3}$. This is analogous
to the principle employed in \citet{2017ApJ...844..161A} to increase FRB
discovery rates with the CHIME Pathfinder using
an ``incoherent formed beam''.
}

\refchange{We can use Equation~\ref{e:wA_from_wdelta} to form the same beam from the
FFT beams $b_A$ instead of the $\widetilde{V}_\delta$. This gives:}
\begin{align}
    \label{e:wA_windowed}
    w_\textrm{ow}^A
    &=
    \frac{\nant}{M\mathcal{N}}
    \sumdelta
    \frac{%
    e^{i 2 \pi \delta A/M}
    w_p^{\deltaindex}(\theta_p)
    }{(\nant - |\deltaindex|)}
    \sumalpha
    h^{\alphaindex+\delta} h^{\alphaindex *}
    \nonumber
    \\
    &=
    \frac{\nant}{M\mathcal{N}}
    \sum_{\deltaindex\alphaindex}
    \frac{%
    e^{i2\pi\delta[A - (d/\lambda)\sin \theta_p]/M)}
    }{(\nant - |\deltaindex|)}
    h^{\alphaindex+\delta} h^{\alphaindex *}\;.
\end{align}

\refchange{The expression $(\nant - |\deltaindex|)$ is proportional to the
Fourier transform over $A$ of the
function in Equation~\ref{e:bp_max_arbitrary}. As such, by the convolution
theorem, dividing by this expression and the subsequent Fourier-transform-like
operation from $\delta$ to $A$ amount to a \emph{deconvolution} operation on
the window's auto-convolution.}
These coefficients $w^A$ are shown for our example
window in Figure~\ref{f:wA}. We
see that the windowed beam with arbitrary steering angle can be formed with a
much more compact set of weights compared to an unwindowed pointed beam. In the
case shown, a kernel of five weights obtains an excellent approximation to the
full set of weights.
\refchange{As such, in applications where FFT beams are formed in
the initial correlation and then regridded in a post processing step, the
regridding will be computationally more convenient for these optimal windowed
beams.}

A key point is that if we form a full set of $M$ optimal windowed beams, this
is an \emph{invertable} operation on the redundancy stacked visibilities. That
means that the set of optimal windowed beams also contain the same sky
information as the redundancy stacked visibilities.

\begin{figure}
    \begin{center}
        \includegraphics[scale=0.65]{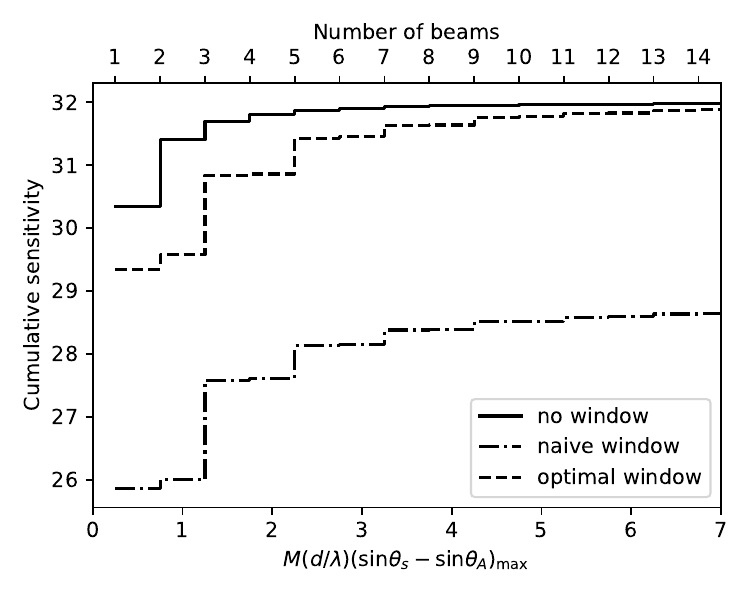}
        \caption{\label{f:cum_info}
        Cumulative sensitivity to an unpolarized point source as a function of 
        the number of
        regularly-spaced beams included. The point source is 
        located $\onequarter$ beam spacing
        from zenith and the cumulative sensitivity is plotted as a
        function of the maximum distance from the source to the steering
        angle, scaled to units of the beam spacing $\lambda/(Md)$.
        We assume the simple noise model in
        Equation~\ref{e:simple_noise}. Sensitivity is relative to an isotropic
        antenna with $M = 2\nant - 1$ and $\nant=32$. We consider \refchange{arrays of
        beams} of the same
        three beam types shown in Figure~\ref{f:windowed_beams}.
        }
    \end{center}
\end{figure}

As previously mentioned the uncertainties in any set of formed beams are, in
general,
correlated.
This obscures somewhat how the sky information is distributed amongst the
beams. For instance, if attempting to measure the flux of a point source not
located at one of the FFT steering angles, it is not clear exactly how that
information is distributed amongst the beams. Figure~\ref{f:wA}
indicates that to form the beam that contains \emph{all} the information, one
needs to take a slowly converging sum over all the FFT formed beams, even while
it is clear from Figure~\ref{f:windowed_beams} that only a small number of
those beams have significant sensitivity to the location of the source.
To get a sense of how the information is distributed amongst the beams,
we calculate the cumulative
sensitivity of a set of beams to an unpolarized point source at angle
$\theta_s$, which in the simple noise model is given by
\begin{equation}
    \textsl{SNR}^2 = \frac{\Delta\nu\Delta t}{T_r^2}
    \sum_{XY} \delta_{\jindex\kindex} B_X^{\jindex\kindex}(\theta_s)
    C^{-1}_{XY} \delta_{lm}
    B_Y^{lm}(\theta_s)\;.
\end{equation}
In Figure~\ref{f:cum_info}, we plot this as a function of the number of beams
included in the sum. We see that while the optimal windowed beams have a more
compact regridding kernels in Figure~\ref{f:wA}, the unwindowed pointed beams
have more compact net information. \refchange{That the unwindowed pointed beams
have more net information than the windowed beams at fixed number is
unsurprising, since the unwindowed beams have a narrower main lobe and are
optimal with no shape constraints, in contrast to the ``optimal windowed
beams.''}

\subsection{Noise correlated between antennas}
\label{s:complicated_noise}

We have so far mostly assumed the simple noise model given in
Equation~\ref{e:simple_noise}. Here we will briefly consider
\refchange{simple,} physically-motivated
departures from this model, what effect they have on sensitivity, and
how the optimal beam forming weights are effected.

One simple case is where the noise continues to dominate the sky,
remains uncorrelated, but the receiver temperature, $T_r$, is feed dependant.
This is expected to result from variations in the properties of the amplifiers
amongst the analogue chains for each feed.
A quick look at
Equation~\ref{e:wopt} shows that the optimal pointed beam can be formed by
weighting the voltages by the inverse receiver temperature, which can
conveniently be done pre-correlation/pre-beamforming.  If this weight is
applied before FFT beamforming, then the sky response of the $b_A$ is modified
but they remain the maximum-sensitivity beams to the same steering angles
$\theta_A$, and they still contain all the sky information.  The Fourier
transform of the $b_A$ becomes a modified version of the $\widetilde{V}_\delta$
where redundant visibilities are co-added with optimal signal-to-noise-square
weights instead of uniform weighting.

Another well motivated noise model includes ``cross talk'': noise coupling
between near-by feeds. Such a model can be written
\begin{equation}
    \label{e:cross_talk_noise}
N_{ab} = T_r \xi_{\delta=a-b}\;,
\end{equation}
where $\xi_{\delta}$ is the noise correlation kernel. We will take $\xi_0=1$,
$\xi_{-\delta} = \xi_{\delta}^*$ and assume that it is compact: that the
correlations are negligible except for $|\delta| \ll \nant$.
Under the assumption of redundancy,
for each noise contribution that couples from antenna $a$ to antenna $b$, there
should be an equal contribution that couples from $b$ to $a$ with the opposite
phase. As such, the
imaginary part of $\xi_{a-b}$ should be zero, although we present the more
general case.
$N_{ab}$ can be inverted analytically if we approximate the array as
being periodic,
allowing us to take its inverse in the spatial Fourier domain.
Define
\begin{equation}
    \xi(\theta) \equiv \sum_{\delta=0}^{\nant - 1}
    e^{-i2\pi \delta (d/\lambda) \sin \theta}
    \xi_{\delta}\;.
\end{equation}
Then it can be shown that
\begin{equation}
    N^{-1}_{ab} \approx \frac{1}{\nant T_r}\sum_{A=0}^{\nant-1}
    \frac{e^{i2\pi (a - b)A/\nant}}{\xi(\sin\theta = A\lambda/(\nant d))}
    \;.
\end{equation}
The approximation improves as $\xi_{\delta}$ becomes more compact compared to
$\nant$, since edge
effects from the assumed periodicity become less significant.
From this, it can be shown that the optimal
weights given in Equation~\ref{e:wopt} are just the normal pointed beam
weights, $w_p^{ab}(\theta_p)$ with no modifications.

However, the variance of these beams gets modified by the correlations
\begin{equation}
    \Var[b_p(\theta_p)] = \frac{T_r^2}{\Delta \nu \Delta t}
    {\xi(\theta_p) \xi(\theta_p)^*}
    \;.
\end{equation}
As such, cross talk induces sky directions of lower sensitivity. Because
$\xi_\delta$ \refchange{is typically real-valued}, the loss of sensitivity
will be strongest in the zenith direction.

Note that sky contributions to the total covariance have a similar effect as
cross-talk, since for a redundant array $S_{ab}$ also only depends on $a - b$.
In this analogue, $T_r \xi(\theta) \rightsquigarrow I(\theta)$.
However, there is no reason to think $S_{ab}$ will be especially compact in
$a-b$, and as such it is unclear if our analytic matrix inversion is at all
valid.

\subsection{Localization}
\label{s:localization}

One common use of multi-beam systems is to determine the sky location of a
source detected in one or more beams. This is especially true in searches for
fast radio bursts where follow-up of the transients are usually impossible. Here
we will derive the optimal maximum-likelihood estimator for the sky
location for the case where the
formed beams are the $M$ FFT beams of a redundant array $b_A$.
One place where this is particularly useful is in triggered baseband recoding
systems, where we
have the freedom to correlate the data in any way we see fit but FFT
beamforming can be done efficiently. We will briefly discuss general sets of
formed beams at the end of the Section.

Define $b^s_X$ as the contribution to beam $b_X$ from a point source,
which we assume can be
cleanly separated from backgrounds (e.g.~in the time domain for transients or
radio spectrum for lines).
Combining Equations~\ref{eq:Jy_to_K} and~\ref{e:response_def}, we have
\begin{align}
    \langle b^s_X \rangle 
    &= \delta_{\jindex\kindex}
    B^{\jindex\kindex}_X(\vnhat_s) \frac{c^2}{2 k_B \nu^2} S^s_\nu
    \nonumber\\
    &=
    w^{ab}_X
    e^{i 2\pi \vu_{ab} \cdot\vnhat_s}
    A_a^\jindex(\vnhat_s, \nu) A_b^\kindex(\vnhat_s, \nu)^*
    \delta_{\jindex\kindex}\frac{c^2}{2 k_B \nu^2} S^s_\nu
    \;.
\end{align}
Or goal is to estimate $\vnhat_s$, noting that there is a second unknown
parameter, the
flux $S^s_\nu$, with which the location may be degenerate.

The log likelihood is
\begin{align}
    \ln \mathcal{L} &= -\frac{1}{2} \chi^2
    \nonumber\\
    &= -\frac{1}{2}\sum_{XY}
    [b^s_Y - \langle b^s_Y(\vnhat_s, S^s_\nu)\rangle]
    \mathcal{C}^{-1}_{YX}
    [b^s_X - \langle b^s_X(\vnhat_s, S^s_\nu)\rangle]
    \;,
    \label{e:likelihood}
\end{align}
where $\mathcal{C}_{XY}$ should be estimated using
a sky and noise model. Alternatively it could be estimated directly from the
data using Equation~\ref{e:beam_cov} should the visibilities---or in redundant
arrays $\tilde{V}_\delta$ or $b_A$---be available.
We would like to find the value of $\vnhat_s$ and $S_\nu^s$ that maximizes this
likelihood (minimizes $\chi^2$).

From here we restrict ourselves to the case where the beams are the
of FFT beams in a redundant array and to the simplified noise model.
We define
$T^s \equiv A^\jindex A^{\kindex*} \delta_{\jindex\kindex}\frac{c^2}{2 k_B \nu^2} S^s_\nu$ and use this
rather than the flux to parameterize the source strength. Note that while
the primary beam sky response, $A^i$ depends on the unknown source location,
this dependence is assumed to be weak compared to the interferometric phases.
\refchange{The validity of this assumption will depend on the instrument's
antenna response and array configuration. For CHIME this is likely an excellent
approximation in the North--South directions but may be invalid East--West.}
As such we will ignore the small amount of information contained in this
dependence.
For our derivation we will initially work with the $\tilde{V}_\delta$ rather
than the $b_A$ since they are uncorrelated in the simple noise model.
\refchange{Thus,
Equation~\ref{e:simple_noise_cov} (scaled by the stacking factor $\nant -
|\delta|$) can be used for the covariance}.
We then have
\begin{align}
    \chi^2
    &=
    \frac{\Delta\nu\Delta t}{T_r^2}
    \sum_{\delta}
    \frac{%
    [V^s_\delta - \langle V^s_\delta(\vnhat_s)\rangle]
    [V^s_\delta - \langle V^s_\delta(\vnhat_s)\rangle]^*
    }{\nant - |\delta|}
    \nonumber\\
    &=
    \frac{\Delta\nu\Delta t}{T_r^2}
    \sum_{\delta}
    \frac{%
        \left|
            V^s_\delta
            - (\nant - |\delta|)T^s e^{i2\pi \delta (d/\lambda) \sin\theta_s}
        \right|^2
    }{\nant - |\delta|}
    \;.
\end{align}

We will use Newton's method to find the minimum. This requires
the first and second derivatives of $\chi^2$ with respect to $\sin\theta_s$.
These are
\begin{align}
    \frac{\partial \chi^2}{\partial \sin\theta_s}
    &=
    \frac{2\Delta\nu\Delta t T_s}{T_r^2}
    \frac{d}{\lambda}
    \sum_{\delta}
    (i2\pi \delta)
    V^{s}_\delta e^{-i2\pi (d/\lambda) \delta \sin\theta_s}
    \\
    \frac{\partial^2 \chi^2}{(\partial \sin\theta_s)^2}
    &=
    \frac{2\Delta\nu\Delta t T_s}{T_r^2}
    \frac{d^2}{\lambda^2}
    \sum_{\delta}
    (2\pi \delta)^2 
    V^{s}_\delta e^{-i2\pi (d/\lambda) \delta \sin\theta_s}
    \;.
\end{align}

The Newton's method estimator for $\sin \theta_s$, which we denote as
$\widehat{\sin \theta_s}$, is then
\begin{align}
    \widehat{\sin \theta_s}
    &= \sin \theta_s
    - \frac{\partial \chi^2}{\partial \sin\theta_s}
    \left[\frac{\partial^2 \chi^2}{(\partial \sin\theta_s)^2}\right]^{-1}
    \nonumber\\
    &= \sin \theta_s
    -
    \frac{\lambda}{d}
    \sum_{\delta}
    \delta i2\pi V^{s}_\delta e^{-i2\pi (d/\lambda) \delta \sin\theta_s}
    \nonumber\\
    &\qquad\times
    \left[
    \sum_{\delta}
    (2\pi \delta)^2 V^{s}_\delta e^{-i2\pi (d/\lambda) \delta \sin\theta_s}
    \right]^{-1}
    \;,
\end{align}
where $\sin\theta_s$ is to be evaluated at the current best guess for the
location and the estimator should be applied iteratively until it converges.
Note that $T_s$ cancels, so there is no
degeneracy with the source flux (except from the primary beam which we have
explicitly ignored). This should generically be true any time the complete
array information, either in the form of $V_{ab}$, $\tilde{V}_\delta$, or
$b_A$, is available.
Inspecting the above formula gives some insight into how the estimator operates.
The first factor of the update term tells us to form the beam whose
sky response is the
derivative of the pointed beam with respect to steering angle $\theta_s$.
The factor in square brackets tells us to form
the beam whose sky response is the curvature (second derivative) of the pointed
beam with steering angle $\theta_s$.

Armed with this form, we can
proceed to make improvements to the simple Newton's-method estimator. First,
Newton's method assumes that the curvature is constant, or at least slowly
varying between the initial guess and the true maximum. Inspecting the beam
shape for the pointed beam in Figure~\ref{f:beams}, we see that there
is actually an inflection point roughly a quarter beam width from the maximum.
Thus, we are almost certainly better off replacing the curvature at the initial
guess with the curvature at the maximum, properly scaled for the best estimate
of the flux. That is
\begin{align}
    &\left[
    \sum_{\delta}
    4\pi^2\delta^2 V^{s}_\delta e^{-i2\pi (d/\lambda) \delta \sin\theta_s}
    \right]
    \nonumber\\
    &\qquad\rightarrow
    \left[
    \sum_{\delta}
    \delta^2 (\nant - |\delta|)
    \right] 4\pi^2\frac{b_p^s(\theta_s)}{\nant}
    \;.
\end{align}
Note that these become equivalent as the estimate converges to the true
maximum. The finite sum in the square brackets is
\begin{align}
    \left[
    \sum_{\delta}
    \delta^2 (\nant - |\delta|)
    \right]
    \nonumber
    &=
    2 \nant^2 (\nant -1)
    \left[
        \frac{2\nant - 1}{6} - \frac{\nant - 1}{4}
    \right]
    \\
    &\approx \frac{\nant^4}{6}
    \;.
    \label{e:baseline_MOI}
\end{align}

Another concern is that if the initial guess is very poor, specifically if it is
off by more than $\lambda/(\nant d)$, the method will converge to a local maximum
in the likelihood associated with a sidelobe of the pointed beam rather than the
global maximum. This should rarely be a problem, since the FFT formed beams are
spaced by $\lambda/(M d)$ with $M \geq 2\nant -1$, so an inital guess based on
the beam of strongest detection should be off by less than half this.

Making the above substitutions and rewriting in terms of $b_A$, we have
\begin{align}
    \widehat{\sin \theta_s}
    &= \sin \theta_s
    -
    \frac{\lambda}{d}
    \frac{\nant}{M}
    \sum_{\delta A}
    \delta i2\pi e^{-i2\pi \delta \left[(d/\lambda) \sin\theta_s - A/M\right]}
    b^s_A
    \nonumber\\
    &\qquad\qquad\times
    \left[
    \frac{4\pi^2 \nant^3 b_p^s(\theta_s)}{6}
    \right]^{-1}
    \nonumber
    \\
    &= \sin \theta_s
    +
    \frac{\lambda}{d}
    \sum_A \left. \frac{d w^A_p}{d y_p^A} \right|_{\theta_p = \theta_s}
    b^s_A
    \nonumber\\
    &\qquad\qquad\times
    \left[
    \frac{4\pi^2\nant^3}{6}
    \sum_B w^B_p(\theta_s) b^s_B
    \right]^{-1}\;,
\end{align}
where $w^A_p$ is given in Equation~\ref{e:bp_max_arbitrary} in terms
of $y_p^A$. Finally, it
is clear from Figure~\ref{f:cum_info} that the information about a point source
is nearly completely confined to the $\sim10$ FFT formed beams surrounding its
location. As such, for applications where computational efficiency is important, we
recommend truncating the sums over $A$ and $B$ in the above formulae to only
include those beams. If the beams retained contain
$\sim 99\%$ of the information about the point source, this will have a
negligible effect on the convergence of the estimator and the final location
uncertainty.

The statistical uncertainty of this estimator, once converged, can be written
in terms of the curvature of the likelihood at its maximum, which in turn can
be written in terms of the total signal-to-noise ratio $\textit{SNR}$:
\begin{align}
    \Var(\widehat{\sin\theta_s})
    &= \left[\frac{1}{2} 
    \left\langle\frac{\partial^2 \chi^2}{(\partial \sin\theta_s)^2}\right\rangle
    \right]^{-1}
    \nonumber\\
    &=
    \left(
        \frac{\sqrt{6}}{2\pi}
        \frac{\lambda}{\nant d}
        \frac{1}{\textit{SNR}}
    \right)^2
    \;.
\end{align}
The factor of $\sqrt 6$ in the above equation comes from
Equation~\ref{e:baseline_MOI} and is related to the moment of inertia of the
baseline distribution for a linear array.

In the general case where the beam $b^s_X$ are not the FFT beams (or
$\widetilde V_\delta$, or an equivalent full-information set), an
analogous estimator could be derived following the same procedure. One concern
is that the Newton's method estimator might converge to a local maximum of the
likelihood, which we have argued is not a concern for the FFT beams.
Alternately, the likelihood in Equation~\ref{e:likelihood} could be sampled using
Markov-Chain Monte Carlo or other sampling techniques.


\section{Discussion and Conclusions}

We have developed a formalism for describing beamforming in radio
interferometers. One aspect that distinguishes our formalism from previous
work is that we do not ignore correlations in the visibility-space noise
due to the sky or noise coupling between receivers. These correlations
are more relevant at low frequencies, where the sky signal typically
dominates noise, and for compact interferometers where noise coupling is
stronger. These are exactly the regimes where FFT beamforming will be most
used, and so future analyses will need to consider these correlations.
We have shown in Section~\ref{s:fft_beamforming}
that FFT beamforming---which is equivalent
in its information
content to redundant visibility stacking---does not strictly preserve the sky
information content of the full $n^2$ correlation. This is because both the
sky and noise coupling induce noise correlations between the visibilities
that depend upon which antennas participate in the visibilities. This is true
even amongst visibilities that are perfectly redundant, in that they have
the same expectation value from both sky and noise. As such, a simple stack
of redundant visibilities that ignores these correlations is sub-optimal.

The simplest implementations of FFT beamforming give very limited control
over the location of the formed beams. As such, when performing targeted
observations of point sources, instruments typically form tied-array beams
from the digitized voltages, rather than FFT beams, even when the number of
tied-array beams is large (\textit{eg.} \cite{2017MNRAS.468.3746C}). However,
Equation~\ref{e:bp_max_arbitrary} permits arbitrary tied-array beams to be
formed from the FFT beams in intensity rather than voltage. Assuming that
intensity-space operations are far cheaper than
voltage-space operations---\textsl{i.e.}~the downsampling
$\Delta \nu \Delta t \gg 1$---this procedure is
computationally cheaper as long as the number of output beams is larger
than $\sim \log \nant$.

Sculpting the beam shape, for example to suppress sidelobes, is also more
effectively done on FFT beam intensity rather than in voltage space. 
\refchange{In contrast to the normal procedure of applying a spatial window
function to the antenna signals prior to beamforming, forming a sculpted beam
from the FFT beams preserves more sky information for the same beam shape.}
Windowed beams can also be formed to
arbitrary steering angles from a
more compact subset of the FFT
beams, as shown in Figure~\ref{f:wA}, which may have advantages in
applications where the computational cost of this operation is significant.

When localizing a source (such as a fast radio burst) from multi-beam
detections, it is important to account for noise correlations in the formed
beams, since these are present even in cases where the visibilities are
uncorrelated.  Failure to account for these correlations will result in a
sub-optimal estimator and/or mis-estimations of the localization uncertainty.
We have derived a source-location estimator that operates directly on the
FFT beams, and accounts for these correlations for simple noise models.
This estimator may be particularly useful in triggered baseband recording
systems, where there is complete freedom in how to correlate the data.

Radio astronomy is expected to become increasingly reliant on FFT
beamforming as the scale of instruments grows. As such, it will be
increasingly important to have an understanding of the capabilities and
limitations of the algorithm such that the substantial potential of
upcoming instruments can be realized.

\acknowledgements

During this work K.W.M was supported in part by the Canadian Institute for Theoretical Astrophysics
National Fellows program.

\bibliography{beamforming.bib}

\begin{thebibliography}{}
\expandafter\ifx\csname natexlab\endcsname\relax\def\natexlab#1{#1}\fi

\bibitem[{{Amiri} {et~al.}(2017){Amiri}, {Bandura}, {Berger}, {Bond}, {Cliche},
  {Connor}, {Deng}, {Denman}, {Dobbs}, {Domagalski}, {Fandino}, {Gilbert},
  {Good}, {Halpern}, {Hanna}, {Hincks}, {Hinshaw}, {H{\"o}fer}, {Hsyu},
  {Klages}, {Landecker}, {Masui}, {Mena-Parra}, {Newburgh}, {Oppermann}, {Pen},
  {Peterson}, {Pinsonneault-Marotte}, {Renard}, {Shaw}, {Siegel}, {Sigurdson},
  {Smith}, {Storer}, {Tretyakov}, {Vanderlinde}, \&
  {Wiebe}}]{2017ApJ...844..161A}
{Amiri}, M., {Bandura}, K., {Berger}, P., {et~al.} 2017, \apj, 844, 161

\bibitem[{{Bandura} {et~al.}(2014){Bandura}, {Addison}, {Amiri}, {Bond},
  {Campbell-Wilson}, {Connor}, {Cliche}, {Davis}, {Deng}, {Denman}, {Dobbs},
  {Fandino}, {Gibbs}, {Gilbert}, {Halpern}, {Hanna}, {Hincks}, {Hinshaw},
  {H{\"o}fer}, {Klages}, {Landecker}, {Masui}, {Mena Parra}, {Newburgh}, {Pen},
  {Peterson}, {Recnik}, {Shaw}, {Sigurdson}, {Sitwell}, {Smecher}, {Smegal},
  {Vanderlinde}, \& {Wiebe}}]{2014SPIE.9145E..22B}
{Bandura}, K., {Addison}, G.~E., {Amiri}, M., {et~al.} 2014, in \procspie, Vol.
  9145, Ground-based and Airborne Telescopes V, 914522

\bibitem[{{Beardsley} {et~al.}(2017){Beardsley}, {Thyagarajan}, {Bowman}, \&
  {Morales}}]{2017MNRAS.470.4720B}
{Beardsley}, A.~P., {Thyagarajan}, N., {Bowman}, J.~D., \& {Morales}, M.~F.
  2017, \mnras, 470, 4720

\bibitem[{{Berger} {et~al.}(2016){Berger}, {Newburgh}, {Amiri}, {Bandura},
  {Cliche}, {Connor}, {Deng}, {Denman}, {Dobbs}, {Fandino}, {Gilbert}, {Good},
  {Halpern}, {Hanna}, {Hincks}, {Hinshaw}, {H{\"o}fer}, {Johnson}, {Landecker},
  {Masui}, {Mena Parra}, {Oppermann}, {Pen}, {Peterson}, {Recnik}, {Robishaw},
  {Shaw}, {Siegel}, {Sigurdson}, {Smith}, {Storer}, {Tretyakov}, {Van Gassen},
  {Vanderlinde}, \& {Wiebe}}]{2016SPIE.9906E..0DB}
{Berger}, P., {Newburgh}, L.~B., {Amiri}, M., {et~al.} 2016, in \procspie, Vol.
  9906, Ground-based and Airborne Telescopes VI, 99060D

\bibitem[{{Caleb} {et~al.}(2016){Caleb}, {Flynn}, {Bailes}, {Barr}, {Bateman},
  {Bhandari}, {Campbell-Wilson}, {Green}, {Hunstead}, {Jameson}, {Jankowski},
  {Keane}, {Ravi}, {van Straten}, \& {Krishnan}}]{2016MNRAS.458..718C}
{Caleb}, M., {Flynn}, C., {Bailes}, M., {et~al.} 2016, \mnras, 458, 718

\bibitem[{{Caleb} {et~al.}(2017){Caleb}, {Flynn}, {Bailes}, {Barr}, {Bateman},
  {Bhandari}, {Campbell-Wilson}, {Farah}, {Green}, {Hunstead}, {Jameson},
  {Jankowski}, {Keane}, {Parthasarathy}, {Ravi}, {Rosado}, {van Straten}, \&
  {Venkatraman Krishnan}}]{2017MNRAS.468.3746C}
---. 2017, \mnras, 468, 3746

\bibitem[{{Chen}(2012)}]{2012IJMPS..12..256C}
{Chen}, X. 2012, in International Journal of Modern Physics Conference Series,
  Vol.~12, International Journal of Modern Physics Conference Series, 256--263

\bibitem[{{CHIME/FRB Collaboration} {et~al.}(2018){CHIME/FRB Collaboration},
  {Amiri}, {Bandura}, {Berger}, {Bhardwaj}, {Boyce}, {Boyle}, {Brar},
  {Burhanpurkar}, {Chawla}, {Chowdhury}, {Cliche}, {Cranmer}, {Cubranic},
  {Deng}, {Denman}, {Dobbs}, {Fandino}, {Fonseca}, {Gaensler}, {Giri},
  {Gilbert}, {Good}, {Guliani}, {Halpern}, {Hinshaw}, {H{\"o}fer}, {Josephy},
  {Kaspi}, {Landecker}, {Lang}, {Liao}, {Masui}, {Mena-Parra}, {Naidu},
  {Newburgh}, {Ng}, {Patel}, {Pen}, {Pinsonneault-Marotte}, {Pleunis}, {Rafiei
  Ravandi}, {Ransom}, {Renard}, {Scholz}, {Sigurdson}, {Siegel}, {Smith},
  {Stairs}, {Tendulkar}, {Vanderlinde}, \& {Wiebe}}]{2018ApJ...863...48C}
{CHIME/FRB Collaboration}, {Amiri}, M., {Bandura}, K., {et~al.} 2018, \apj,
  863, 48

\bibitem[{Daishido {et~al.}(2000)Daishido, Tanaka, Takeuchi, Akamine, Fujii,
  Kuniyoshi, Suemitsu, Gotoh, Mizuki, Mizuno, Suzuki, \& Asuma}]{daishido2000}
Daishido, T., Tanaka, N., Takeuchi, H., {et~al.} 2000, in \procspie, Vol. 4015,
  Radio Telescopes

\bibitem[{{DeBoer} {et~al.}(2017){DeBoer}, {Parsons}, {Aguirre}, {Alexander},
  {Ali}, {Beardsley}, {Bernardi}, {Bowman}, {Bradley}, {Carilli}, {Cheng}, {de
  Lera Acedo}, {Dillon}, {Ewall-Wice}, {Fadana}, {Fagnoni}, {Fritz},
  {Furlanetto}, {Glendenning}, {Greig}, {Grobbelaar}, {Hazelton}, {Hewitt},
  {Hickish}, {Jacobs}, {Julius}, {Kariseb}, {Kohn}, {Lekalake}, {Liu}, {Loots},
  {MacMahon}, {Malan}, {Malgas}, {Maree}, {Martinot}, {Mathison}, {Matsetela},
  {Mesinger}, {Morales}, {Neben}, {Patra}, {Pieterse}, {Pober}, {Razavi-Ghods},
  {Ringuette}, {Robnett}, {Rosie}, {Sell}, {Smith}, {Syce}, {Tegmark},
  {Thyagarajan}, {Williams}, \& {Zheng}}]{2017PASP..129d5001D}
{DeBoer}, D.~R., {Parsons}, A.~R., {Aguirre}, J.~E., {et~al.} 2017, \pasp, 129,
  045001

\bibitem[{{Foster} {et~al.}(2014){Foster}, {Hickish}, {Magro}, {Price}, \&
  {Zarb Adami}}]{2014MNRAS.439.3180F}
{Foster}, G., {Hickish}, J., {Magro}, A., {Price}, D., \& {Zarb Adami}, K.
  2014, \mnras, 439, 3180

\bibitem[{{Furlanetto} {et~al.}(2006){Furlanetto}, {Oh}, \&
  {Briggs}}]{2006PhR...433..181F}
{Furlanetto}, S.~R., {Oh}, S.~P., \& {Briggs}, F.~H. 2006, \physrep, 433, 181

\bibitem[{{Greenhill} {et~al.}(2012){Greenhill}, {Werthimer}, {Taylor},
  {Ellingson}, \& {LEDA Collaboration}}]{2012AAS...22010403G}
{Greenhill}, L.~J., {Werthimer}, D., {Taylor}, G., {Ellingson}, S., \& {LEDA
  Collaboration}. 2012, in American Astronomical Society Meeting Abstracts,
  Vol. 220, American Astronomical Society Meeting Abstracts \#220, 104.03

\bibitem[{{Hamaker} {et~al.}(1996){Hamaker}, {Bregman}, \&
  {Sault}}]{1996A&AS..117..137H}
{Hamaker}, J.~P., {Bregman}, J.~D., \& {Sault}, R.~J. 1996, \aaps, 117, 137

\bibitem[{IEEE(2014)}]{6758443}
IEEE. 2014, IEEE Std 145-2013 (Revision of IEEE Std 145-1993), 1

\bibitem[{{Kulkarni}(1989)}]{1989AJ.....98.1112K}
{Kulkarni}, S.~R. 1989, \aj, 98, 1112

\bibitem[{{Liu} {et~al.}(2010){Liu}, {Tegmark}, {Morrison}, {Lutomirski}, \&
  {Zaldarriaga}}]{2010MNRAS.408.1029L}
{Liu}, A., {Tegmark}, M., {Morrison}, S., {Lutomirski}, A., \& {Zaldarriaga},
  M. 2010, \mnras, 408, 1029

\bibitem[{{Liu} {et~al.}(2009){Liu}, {Tegmark}, \&
  {Zaldarriaga}}]{2009MNRAS.394.1575L}
{Liu}, A., {Tegmark}, M., \& {Zaldarriaga}, M. 2009, \mnras, 394, 1575

\bibitem[{{Loeb} \& {Zaldarriaga}(2004)}]{2004PhRvL..92u1301L}
{Loeb}, A., \& {Zaldarriaga}, M. 2004, Physical Review Letters, 92, 211301

\bibitem[{{Lonsdale} {et~al.}(2009){Lonsdale}, {Cappallo}, {Morales}, {Briggs},
  {Benkevitch}, {Bowman}, {Bunton}, {Burns}, {Corey}, {Desouza}, {Doeleman},
  {Derome}, {Deshpande}, {Gopala}, {Greenhill}, {Herne}, {Hewitt}, {Kamini},
  {Kasper}, {Kincaid}, {Kocz}, {Kowald}, {Kratzenberg}, {Kumar}, {Lynch},
  {Madhavi}, {Matejek}, {Mitchell}, {Morgan}, {Oberoi}, {Ord},
  {Pathikulangara}, {Prabu}, {Rogers}, {Roshi}, {Salah}, {Sault}, {Shankar},
  {Srivani}, {Stevens}, {Tingay}, {Vaccarella}, {Waterson}, {Wayth}, {Webster},
  {Whitney}, {Williams}, \& {Williams}}]{2009IEEEP..97.1497L}
{Lonsdale}, C.~J., {Cappallo}, R.~J., {Morales}, M.~F., {et~al.} 2009, IEEE
  Proceedings, 97, 1497

\bibitem[{{Masui} {et~al.}(2015){Masui}, {Amiri}, {Connor}, {Deng}, {Fandino},
  {H{\"o}fer}, {Halpern}, {Hanna}, {Hincks}, {Hinshaw}, {Parra}, {Newburgh},
  {Shaw}, \& {Vanderlinde}}]{2015A&C....12..181M}
{Masui}, K., {Amiri}, M., {Connor}, L., {et~al.} 2015, Astronomy and Computing,
  12, 181

\bibitem[{{Masui} \& {Pen}(2010)}]{2010PhRvL.105p1302M}
{Masui}, K.~W., \& {Pen}, U.-L. 2010, Physical Review Letters, 105, 161302

\bibitem[{{Morales}(2011)}]{2011PASP..123.1265M}
{Morales}, M.~F. 2011, \pasp, 123, 1265

\bibitem[{{Nakajima} {et~al.}(1992){Nakajima}, {Otobe}, {Nishiboru},
  {Watanabe}, {Asuma}, \& {Daishido}}]{1992PASJ...44L..35N}
{Nakajima}, J., {Otobe}, E., {Nishiboru}, K., {et~al.} 1992, \pasj, 44, L35

\bibitem[{{Newburgh} {et~al.}(2014){Newburgh}, {Addison}, {Amiri}, {Bandura},
  {Bond}, {Connor}, {Cliche}, {Davis}, {Deng}, {Denman}, {Dobbs}, {Fandino},
  {Fong}, {Gibbs}, {Gilbert}, {Griffin}, {Halpern}, {Hanna}, {Hincks},
  {Hinshaw}, {H{\"o}fer}, {Klages}, {Landecker}, {Masui}, {Parra}, {Pen},
  {Peterson}, {Recnik}, {Shaw}, {Sigurdson}, {Sitwell}, {Smecher}, {Smegal},
  {Vanderlinde}, \& {Wiebe}}]{2014SPIE.9145E..4VN}
{Newburgh}, L.~B., {Addison}, G.~E., {Amiri}, M., {et~al.} 2014, in \procspie,
  Vol. 9145, Ground-based and Airborne Telescopes V, 91454V

\bibitem[{{Newburgh} {et~al.}(2016){Newburgh}, {Bandura}, {Bucher}, {Chang},
  {Chiang}, {Cliche}, {Dav{\'e}}, {Dobbs}, {Clarkson}, {Ganga}, {Gogo},
  {Gumba}, {Gupta}, {Hilton}, {Johnstone}, {Karastergiou}, {Kunz}, {Lokhorst},
  {Maartens}, {Macpherson}, {Mdlalose}, {Moodley}, {Ngwenya}, {Parra},
  {Peterson}, {Recnik}, {Saliwanchik}, {Santos}, {Sievers}, {Smirnov},
  {Stronkhorst}, {Taylor}, {Vanderlinde}, {Van Vuuren}, {Weltman}, \&
  {Witzemann}}]{2016SPIE.9906E..5XN}
{Newburgh}, L.~B., {Bandura}, K., {Bucher}, M.~A., {et~al.} 2016, in \procspie,
  Vol. 9906, Ground-based and Airborne Telescopes VI, 99065X

\bibitem[{{Ng} {et~al.}(2017){Ng}, {Vanderlinde}, {Paradise}, {Klages},
  {Masui}, {Smith}, {Bandura}, {Boyle}, {Dobbs}, {Kaspi}, {Renard}, {Shaw},
  {Stairs}, \& {Tretyakov}}]{2017arXiv170204728N}
{Ng}, C., {Vanderlinde}, K., {Paradise}, A., {et~al.} 2017, ArXiv e-prints,
  arXiv:1702.04728

\bibitem[{{Otobe} {et~al.}(1994){Otobe}, {Nakajima}, {Nishibori}, {Saito},
  {Kobayashi}, {Tanaka}, {Watanabe}, {Aramaki}, {Hoshikawa}, {Asuma}, \&
  {Daishido}}]{1994PASJ...46..503O}
{Otobe}, E., {Nakajima}, J., {Nishibori}, K., {et~al.} 1994, \pasj, 46, 503

\bibitem[{{Pen}(2004)}]{2004NewA....9..417P}
{Pen}, U.-L. 2004, \na, 9, 417

\bibitem[{{Price} {et~al.}(2017){Price}, {Greenhill}, {Fialkov}, {Bernardi},
  {Garsden}, {Barsdell}, {Kocz}, {Anderson}, {Bourke}, {Craig}, {Dexter},
  {Dowell}, {Eastwood}, {Eftekhari}, {Ellingson}, {Hallinan}, {Hartman},
  {Kimberk}, {Lazio}, {Leiker}, {MacMahon}, {Monroe}, {Schinzel}, {Taylor},
  {Werthimer}, \& {Woody}}]{2017arXiv170909313P}
{Price}, D.~C., {Greenhill}, L.~J., {Fialkov}, A., {et~al.} 2017, ArXiv
  e-prints, arXiv:1709.09313

\bibitem[{{Saiyad Ali} \& {Bharadwaj}(2013)}]{2013arXiv1310.1707S}
{Saiyad Ali}, S., \& {Bharadwaj}, S. 2013, ArXiv e-prints, arXiv:1310.1707

\bibitem[{{Shaw} {et~al.}(2014){Shaw}, {Sigurdson}, {Pen}, {Stebbins}, \&
  {Sitwell}}]{2014ApJ...781...57S}
{Shaw}, J.~R., {Sigurdson}, K., {Pen}, U.-L., {Stebbins}, A., \& {Sitwell}, M.
  2014, \apj, 781, 57

\bibitem[{{Shaw} {et~al.}(2015){Shaw}, {Sigurdson}, {Sitwell}, {Stebbins}, \&
  {Pen}}]{2015PhRvD..91h3514S}
{Shaw}, J.~R., {Sigurdson}, K., {Sitwell}, M., {Stebbins}, A., \& {Pen}, U.-L.
  2015, \prd, 91, 083514

\bibitem[{{Smirnov}(2011)}]{2011A&A...527A.106S}
{Smirnov}, O.~M. 2011, \aap, 527, A106

\bibitem[{{Tegmark} \& {Zaldarriaga}(2009)}]{2009PhRvD..79h3530T}
{Tegmark}, M., \& {Zaldarriaga}, M. 2009, \prd, 79, 083530

\bibitem[{{Tegmark} \& {Zaldarriaga}(2010)}]{2010PhRvD..82j3501T}
---. 2010, \prd, 82, 103501

\bibitem[{{Thompson} {et~al.}(2017){Thompson}, {Moran}, \&
  {Swenson}}]{2017isra.book.....T}
{Thompson}, A.~R., {Moran}, J.~M., \& {Swenson}, Jr., G.~W. 2017,
  {Interferometry and Synthesis in Radio Astronomy, 3rd Edition} (Springer
  International Publishing), doi:10.1007/978-3-319-44431-4

\bibitem[{{Thyagarajan} {et~al.}(2017){Thyagarajan}, {Beardsley}, {Bowman}, \&
  {Morales}}]{2017MNRAS.467..715T}
{Thyagarajan}, N., {Beardsley}, A.~P., {Bowman}, J.~D., \& {Morales}, M.~F.
  2017, \mnras, 467, 715

\bibitem[{{van Cittert}(1934)}]{1934Phy.....1..201V}
{van Cittert}, P.~H. 1934, Physica, 1, 201

\bibitem[{{van Haarlem} {et~al.}(2013){van Haarlem}, {Wise}, {Gunst}, {Heald},
  {McKean}, {Hessels}, {de Bruyn}, {Nijboer}, {Swinbank}, {Fallows},
  {Brentjens}, {Nelles}, {Beck}, {Falcke}, {Fender}, {H{\"o}randel},
  {Koopmans}, {Mann}, {Miley}, {R{\"o}ttgering}, {Stappers}, {Wijers},
  {Zaroubi}, {van den Akker}, {Alexov}, {Anderson}, {Anderson}, {van Ardenne},
  {Arts}, {Asgekar}, {Avruch}, {Batejat}, {B{\"a}hren}, {Bell}, {Bell}, {van
  Bemmel}, {Bennema}, {Bentum}, {Bernardi}, {Best}, {B{\^i}rzan}, {Bonafede},
  {Boonstra}, {Braun}, {Bregman}, {Breitling}, {van de Brink}, {Broderick},
  {Broekema}, {Brouw}, {Br{\"u}ggen}, {Butcher}, {van Cappellen}, {Ciardi},
  {Coenen}, {Conway}, {Coolen}, {Corstanje}, {Damstra}, {Davies}, {Deller},
  {Dettmar}, {van Diepen}, {Dijkstra}, {Donker}, {Doorduin}, {Dromer}, {Drost},
  {van Duin}, {Eisl{\"o}ffel}, {van Enst}, {Ferrari}, {Frieswijk}, {Gankema},
  {Garrett}, {de Gasperin}, {Gerbers}, {de Geus}, {Grie{\ss}meier}, {Grit},
  {Gruppen}, {Hamaker}, {Hassall}, {Hoeft}, {Holties}, {Horneffer}, {van der
  Horst}, {van Houwelingen}, {Huijgen}, {Iacobelli}, {Intema}, {Jackson},
  {Jelic}, {de Jong}, {Juette}, {Kant}, {Karastergiou}, {Koers}, {Kollen},
  {Kondratiev}, {Kooistra}, {Koopman}, {Koster}, {Kuniyoshi}, {Kramer},
  {Kuper}, {Lambropoulos}, {Law}, {van Leeuwen}, {Lemaitre}, {Loose}, {Maat},
  {Macario}, {Markoff}, {Masters}, {McFadden}, {McKay-Bukowski}, {Meijering},
  {Meulman}, {Mevius}, {Middelberg}, {Millenaar}, {Miller-Jones}, {Mohan},
  {Mol}, {Morawietz}, {Morganti}, {Mulcahy}, {Mulder}, {Munk}, {Nieuwenhuis},
  {van Nieuwpoort}, {Noordam}, {Norden}, {Noutsos}, {Offringa}, {Olofsson},
  {Omar}, {Orr{\'u}}, {Overeem}, {Paas}, {Pandey-Pommier}, {Pandey}, {Pizzo},
  {Polatidis}, {Rafferty}, {Rawlings}, {Reich}, {de Reijer}, {Reitsma},
  {Renting}, {Riemers}, {Rol}, {Romein}, {Roosjen}, {Ruiter}, {Scaife}, {van
  der Schaaf}, {Scheers}, {Schellart}, {Schoenmakers}, {Schoonderbeek},
  {Serylak}, {Shulevski}, {Sluman}, {Smirnov}, {Sobey}, {Spreeuw}, {Steinmetz},
  {Sterks}, {Stiepel}, {Stuurwold}, {Tagger}, {Tang}, {Tasse}, {Thomas},
  {Thoudam}, {Toribio}, {van der Tol}, {Usov}, {van Veelen}, {van der Veen},
  {ter Veen}, {Verbiest}, {Vermeulen}, {Vermaas}, {Vocks}, {Vogt}, {de Vos},
  {van der Wal}, {van Weeren}, {Weggemans}, {Weltevrede}, {White}, {Wijnholds},
  {Wilhelmsson}, {Wucknitz}, {Yatawatta}, {Zarka}, {Zensus}, \& {van
  Zwieten}}]{2013A&A...556A...2V}
{van Haarlem}, M.~P., {Wise}, M.~W., {Gunst}, A.~W., {et~al.} 2013, \aap, 556,
  A2

\bibitem[{{Zernike}(1938)}]{1938Phy.....5..785Z}
{Zernike}, F. 1938, Physica, 5, 785

\bibitem[{{Zheng} {et~al.}(2014){Zheng}, {Tegmark}, {Buza}, {Dillon},
  {Gharibyan}, {Hickish}, {Kunz}, {Liu}, {Losh}, {Lutomirski}, {Morrison},
  {Narayanan}, {Perko}, {Rosner}, {Sanchez}, {Schutz}, {Tribiano}, {Valdez},
  {Yang}, {Adami}, {Zelko}, {Zheng}, {Armstrong}, {Bradley}, {Dexter},
  {Ewall-Wice}, {Magro}, {Matejek}, {Morgan}, {Neben}, {Pan}, {Penna},
  {Peterson}, {Su}, {Villasenor}, {Williams}, \& {Zhu}}]{2014MNRAS.445.1084Z}
{Zheng}, H., {Tegmark}, M., {Buza}, V., {et~al.} 2014, \mnras, 445, 1084

\end{thebibliography}



\end{document}